# Air–Sea Interactions on Titan: Effect of Radiative Transfer on the Lake Evaporation and Atmospheric Circulation

Audrey Chatain[1,2], Scot C. R. Rafkin[1], Alejandro Soto[1], Ricardo Hueso[2], and Aymeric Spiga[3]
[1] Department of Space Studies, Southwest Research Institute (SwRI), 1050 Walnut Street, Suite 300, Boulder, CO 80302, USA; audrey.chatain@boulder.swri.edu
[2] Departamento de Física Aplicada, Escuela de Ingeniería de Bilbao, Universidad del País Vasco/Euskal Herriko Unibertsitatea (UPV/EHU), Plaza Ingeniero Torres Quevedo 1, E-48013 Bilbao, Spain
[3] Laboratoire de Météorologie Dynamique/Institut Pierre-Simon Laplace (LMD/IPSL), Centre National de la Recherche Scientifique (CNRS), Sorbonne Université, 4 place Jussieu, Tour 45-55 3$^e$ étage, F-75252 Paris, France

## Abstract

Titan's northern high latitudes host many large hydrocarbon lakes. Like water lakes on Earth, Titan's lakes are constantly subject to evaporation. This process strongly affects the atmospheric methane abundance, the atmospheric temperature, the lake mixed layer temperature, and the local wind circulation. In this work we use a 2D atmospheric mesoscale model coupled to a slab lake model to investigate the effect of solar and infrared radiation on the exchange of energy and methane between Titan's lakes and atmosphere. The magnitude of solar radiation reaching the surface of Titan through its thick atmosphere is only a few watts per square meter. However, we find that this small energy input is important and is comparable in absolute magnitude to the latent and sensible heat fluxes, as suggested in a study by Rafkin & Soto (2020). The implementation of a gray radiative scheme in the model confirms the importance of radiation when studying lakes at the surface of Titan. Solar and infrared radiation change the energy balance of the system leading to an enhancement of the methane evaporation rate, an increase of the equilibrium lake temperature almost completely determined by its environment (humidity, insolation, and background wind), and a strengthening of the local sea breeze, which undergoes diurnal variations. The sea breeze efficiently transports methane vapor horizontally, from the lake to the land, and vertically due to rising motion along the sea breeze front and due to radiation-induced turbulence over the land.

*Unified Astronomy Thesaurus concepts:* Titan (2186); Natural satellite atmospheres (2214); Radiative transfer (1335); Planetary climates (2184); Planetary atmospheres (1244)

## 1. Introduction

Titan is the only place beyond the Earth known to have lakes and seas (Stofan et al. 2007; Hayes 2016). As on Earth where the evaporation of water from oceans, seas, and lakes drives the planet water cycle, the evaporation of methane from lakes and seas on Titan is an important element for the atmospheric circulation (Tokano 2009a) and the most probable source of methane to the atmosphere of Titan (Lunine & Lorenz 2009).

Prior studies have investigated the magnitude of the evaporative processes on Titan through relatively simple analytical models (e.g., Mitri et al. 2007) and more complex mesoscale modeling (e.g., Rafkin & Soto 2020). All of these studies assumed that radiative forcing was unimportant. Indeed, due to its farther distance to the Sun, and its thick and aerosol covered atmosphere, only a few watts per square meter reach the surface of Titan at maximum, compared to several hundreds of watts per square meter in the case of the Earth. Mitri et al. (2007) predicted turbulent sensible and latent heat fluxes at the surface much larger in magnitude than the radiative fluxes. However, Rafkin & Soto (2020) showed that the results from Mitri et al. (2007) were driven by the assumption of a constant air temperature. The lack of atmospheric cooling resulted in large sensible heat fluxes as the lake temperature dropped through evaporative cooling. Thus, the surface fluxes in Mitri et al. (2007) were overestimated compared to a scenario where a cold and moist marine layer could develop over the lake. Rather than having large turbulent fluxes, Rafkin & Soto (2020) indicated that the fluxes trended toward small values, which often approach values close in magnitude to the insolation. Rafkin & Soto (2020) thus concluded that neglecting radiative forcing may not be a good assumption. In the absence of radiative forcing, both Mitri et al. (2007) and Rafkin & Soto (2020) independently found that a local balance of energy was achieved whereby the sensible heat flux—driven by the temperature difference between the lake and atmosphere—was opposite and equal in magnitude to the latent heat flux—driven by the difference between the saturation vapor pressure over the lake and the relative humidity of the atmosphere.

The objective of this study is to investigate whether the addition of radiative forcing upsets the flux balance between lakes and the overlying atmosphere found in prior studies, and if so, in what way. The study uses the same mesoscale model used in Rafkin & Soto (2020) and incorporates a gray radiative transfer scheme (Section 2). A reference simulation with radiation is compared to the case without radiation to identify the mechanisms by which radiative forcing affects the mesoscale circulation over lakes on Titan (Section 3). A sensitivity study is then performed on other parameters of the model to understand how they are affected by radiative forcing (Section 4). The effect of seasonal and latitudinal insolation variations is also quantified (Section 5). We finally discuss the consequences of these new results relatively to current research questions on Titan (Section 6).

## 2. Model Description

### 2.1. Settings and Improvements of the Model

We use in this study the Titan mesoscale model mtWRF previously described in Rafkin & Soto (2020), and configured almost identically. This model is based on the Weather Research and Forecasting (WRF) model and uses the Advanced Research WRF (ARW-WRF) dynamical core (Skamarock et al. 2008). We run 2D simulations because they allow faster performance and more numerous testing of the relevant parameters that we explore here. Limitations include the absence of topography, the impossibility to define complex lake shorelines, and the removal of vorticity-induced effects. Therefore, the results quantify the relative importance of the different phenomena and processes around lakes on Titan under idealized conditions.

The simulation domain is 3200 km wide with 2 km horizontal resolution and 59 atmospheric vertical levels stretched from 3 m at the lowest level to 20 km at the top of the domain. Simulation time is given in Titan days (tsols), with one tsol corresponding to ∼15.9 days on Earth. The center of the domain is occupied by a 300 km-wide lake that is represented by a slab lake model. The slab consists of a single layer of liquid methane with a temperature that instantaneously responds to a net change in energy. The depth of this layer represents the mixing depth of the lake, which is not necessarily the depth of the lake. Mixing depths of 1 m, 10 m, and 100 m are investigated in this paper, as there is evidence of lakes exceeding 100 m depth on Titan from Cassini radar measurements (Mastrogiuseppe et al. 2019). The width of the model lake is comparable to the size of large lakes on Titan: the six largest have a length of 1170 km (Kraken Mare), 500 km (Ligeia Mare), 380 km (Punga Mare), 240 × 90 km (Jingpo Lacus), 220 × 60 km (Ontario Lacus), and 200 km (Hammar Lacus). In addition, Titan also hosts at least 45 lakes with sizes between 30 and 200 km. These smaller lakes should also show similar evaporation processes scaled down to their sizes as explored in Rafkin & Soto (2020). Although the simulation domain length is nonnegligible compared to Titan's circumference of 16,179 km, the size of the domain is mainly chosen to avoid numerical edge effects on the lake-induced circulation. For simplicity, we therefore use a spatially uniform insolation over the domain.

In the simulations presented by Rafkin & Soto (2020), the land surrounding the lake was at a fixed temperature (93.47 K), which was dictated by the Huygens atmospheric temperature profile. Instead of a fixed temperature, we used the WRF soil-slab model described in the appendix of Blackadar (1979), modified for use on Titan, including using a thermal inertia of 600 J m$^{-2}$ K$^{-1}$ s$^{-0.5}$ typical of plains and lakes (MacKenzie et al. 2019b). We highlight that, in this model, the subsurface conduction flux is proportional to the soil thermal inertia. The soil-slab model implements basic physics with a minimum of empirical tuning, and the fundamental physics are captured with sufficient fidelity to match our very limited knowledge of Titan's subsurface. Thus, the surface temperature is free to vary through subsurface conduction, through radiative fluxes, and through sensible heat exchange with the overlying atmosphere. The subsurface soil temperature at an infinitely deep lower boundary is set to a constant. There is currently no measurement of the subsurface temperature, but its value has a significant influence on the surface temperature. Therefore, we also investigate the sensitivity to this parameter in this work. No evaporation, condensation, or adsorption of methane is allowed on the land. The land is dry and the latent heat flux is zero.

The numerical accumulation of errors in the evaluation of prognostic variables (i.e., tendencies) was also improved compared to Rafkin & Soto (2020). Wind, temperature, and methane vapor tendencies are small under Titan's conditions, and their values after one time step were often at or below the numerical precision of an 8 byte float. The addition of an accumulator for the surface temperature tendency in Rafkin & Soto (2020) ameliorated much of this problem. Here, we further improve this technique by increasing the dynamical time step from 15.9 to 159 s to further compensate for small tendencies and to better correct the numerical precision without having to invoke the multi-time-step accumulator as frequently as in Rafkin & Soto (2020). The larger time step produces small, inconsequential changes in the output, mainly during the short spin-up of the simulation. Results with dynamical time steps of 79.5 s and 190.8 s effectively give the same results as 159 s, demonstrating that the numerical precision problem is ameliorated for this range of values.

Finally, a gray radiation scheme based on Schneider at al. (2012) was incorporated. The description of solar scattering in the atmosphere was modified and the short wave radiative transfer parameters were adjusted to better fit the net solar flux profile at the Huygens landing site (Tomasko et al. 2008a). Details on the description, implementation, and tuning of the gray radiative scheme are given in the Appendix. The gray scheme treats broadband solar and broadband infrared energy separately (Weaver & Ramanathan 1995). The solar flux enters the top of the atmosphere, is absorbed and scattered in the atmosphere, and is reflected at the surface. The thermal (infrared) flux is emitted and absorbed by the surface and the atmosphere. All of these computations are done at each dynamical time step. All simulations presented here are started at 00:00 local time (midnight). Initializing the model at different local times has almost no effect on the evolution of the stabilized, diurnally repeatable solution.

Due to all of the above improvements and modifications, the new results should be taken as more realistic and accurate results that supersede the results of Rafkin & Soto (2020). While the general evolution of the system found by Rafkin & Soto (2020) is mirrored in this study, there are important details and new behaviors that arise with the inclusion of radiation and the land surface model.

### 2.2. Initialization

To investigate the sensitivity of the model to various parameters and to explore the effects of diurnal, seasonal, and geographical variations of the solar insolation, several sets of simulations listed in Table 1 were performed. The studied parameters were the initial relative humidity of the lowest atmospheric layer, the subsurface temperature boundary condition, the initial lake surface temperature, the lake mixed layer depth, the background wind speed, the solar longitude (season), and the latitude. Seas and lakes on Titan are mostly found at high latitudes (Punga Mare at 85°N, Ligeia Mare at 79°N, Jingpo Lacus at 73°N, Ontario Lacus at 72°S, and Kraken Mare at 68°N), although some are observed at mid-latitudes (Hammar Lacus at 49°N, Sionascaig Lacus at 42°S, and Urmia Lacus at 39°S; Griffith et al. 2012; Vixie et al. 2015; Tokano 2020). As the scope of this paper is to study the effect of

Table 1
Parameter Settings for All of the Simulations

| Simulation Name | Initial Surface Relative Humidity (%) | Deep Subsurface Temperature (K) | Initial Lake Surface Temperature (K) | Lake Mixed Layer Depth (m) | Background Wind (m s$^{-1}$) | Ls (°) | Latitude (°) |
|---|---|---|---|---|---|---|---|
| S-A[a] | 45 | 93.47 | 90.5 | 1 | 0 | 0 | 42 |
| S-B | 45 | **93.21** | 90.5 | 1 | 0 | 0 | 42 |
| S-C | **0** | 93.47 | **86.5** | 1 | 0 | 0 | 42 |
| S-D | **0** | **93.21** | **86.5** | 1 | 0 | 0 | 42 |
| S-E | **20** | 93.47 | **88** | 1 | 0 | 0 | 42 |
| S-F | **70** | 93.47 | **92** | 1 | 0 | 0 | 42 |
| S-G | 45 | 93.47 | **93.47** | 1 | 0 | 0 | 42 |
| S-H | 45 | **93.21** | **93.47** | 1 | 0 | 0 | 42 |
| S-I | 45 | 93.47 | **88** | 1 | 0 | 0 | 42 |
| S-J | 45 | 93.47 | 90.5 | **10** | 0 | 0 | 42 |
| S-K | 45 | 93.47 | 90.5 | **100** | 0 | 0 | 42 |
| S-L | 45 | 93.47 | 90.5 | 1 | **1** | 0 | 42 |
| S-M | 45 | 93.47 | 90.5 | 1 | **3** | 0 | 42 |
| S-N | 45 | 93.47 | 90.5 | 1 | 0 | **90** | **−85** |
| S-O | 45 | 93.47 | 90.5 | 1 | 0 | **270** | **−42** |
| S-P | 45 | 93.47 | 90.5 | 1 | 0 | **270** | **−85** |
| S-Q[a] | **0** | 93.47 | **93.47** | 1 | 0 | 0 | 42 |
| S-R | 45 | 93.47 | 90.5 | 1 | 0 | **90** | **85** |
| S-S | 45 | 93.47 | 90.5 | 1 | 0 | **90** | **−42** |
| S-T[b] | **20** | 93.47 | **88** | **30** | **1** | **270** | **−72** |
| S-U | 45 | 93.47 | 90.5 | 1 | 0 | 0 | **70** |
| S-V | 45 | 93.47 | 90.5 | 1 | 0 | 0 | **85** |

**Note.** Bold values highlight the differences with the reference simulation S-A. [a] These simulations have been run twice, one with the radiative transfer scheme on, and one with the radiative transfer scheme off. In the text, an "0" is added at the end of the simulation name to signal simulations when the radiation scheme is turned off, e.g., S-A has radiation, and S-A0 does not. [b]S-T is done to be close to Ontario Lacus summer conditions, and it is the only simulation done on a 100 km large lake.

radiation on the evaporation of lakes, and as the solar radiative forcing is higher at lower latitudes, we performed most simulations at the lowest latitude at which lakes are found: 42°. A comparison to higher latitudes is given in Section 5.

The Huygens/HASI temperature profile (Fulchignoni et al. 2005) is used in all of the simulations. The initial land surface temperature was set to the air temperature measured by Huygens at 3 m, and fixed to 93.47 K. This initial condition minimized an initial sensible heat flux due to thermal imbalance between the surface and the atmosphere. We note that on Titan this thermal profile will, in reality, vary with latitude and seasons. But these are only minor variations (Schinder et al. 2012; Newman et al. 2016) that are not expected to greatly affect our investigation, which is focused on the direct effect of radiation on the evaporation of lakes. These are idealized experiments that aim to understand the physical processes at stake and are not intended to exactly reproduce the seasonal and latitudinal thermal conditions for all simulations. All of the results must be interpreted within the context of the idealized experiment assumptions and simplified physics.

Initial methane vapor profiles were computed similarly to the stably stratified profiles described in Rafkin & Soto (2020), taking into account the virtual buoyancy of methane vapor. The methane mixing ratio was kept constant from the surface up to the saturation point, then the saturation curve was followed up to 30 km, and finally the methane mixing ratio remains constant to the model top (see Figure 1). The mixing ratio at the surface is determined by specifying the relative humidity. The definition of the saturation vapor pressure was modified to use a larger range in temperatures compared to Rafkin & Soto (2020), as described in the appendix of Moses et al. (1992).

## 3. The Reference Simulation

In this section, we investigate the effect of radiation on a baseline, reference configuration. Since there are currently little or no measurements of the lake mixed layer depth, the lake temperature or the subsurface temperature, we selected values that are reasonable as a reference case. The sensitivity of the results to other parametric values is discussed in the next section by comparison to the reference case. The lake mixed layer depth is set to 1 m, the initial lake temperature to 90.5 K, and the land underground temperature to 93.47 K (the same as the initial surface temperature in this case). The lake extends 150 km in both directions from the center of the domain, and there is no background wind. The initial relative humidity is set to 45% at the surface. We ran this configuration with the radiative scheme on (simulation S-A, the reference simulation) and with the radiative scheme off (simulation S-A0).

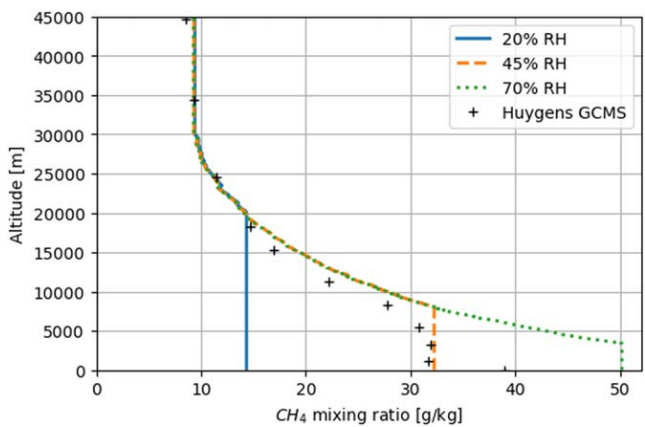

**Figure 1.** Methane mixing ratio profiles used in this work, compared to the Huygens/GCMS measurements (Niemann et al. 2010). Relative humidity (RH in the subpanel) indications are surface values.

The general solution of the reference simulation is similar to that found by Rafkin & Soto (2020). The diurnally stable solution is a sea breeze driven by the temperature difference between the cold air above the lake and the warmer air above the land. However, in our reference simulation, the addition of radiation repartitions the energy budget, and this, in turn, affects the circulation.

### 3.1. Diurnal Variations

The first new result is the appearance of diurnal variation in S-A, which is not present in S-A0 due to its lack of radiative forcing (Figure 2). The diurnal variations in most of the model fields confirm that radiative processes are large enough compared to other energy budget terms to have an impact.

In both S-A and S-A0, a sea breeze, where winds blow from the lake toward the land, forms (Figures 2(a)–(d)). The lake is cooled by evaporation (Figures 2(g) and (h)), and the air above the lake is then cooled by sensible heat flux transfer from the air to the colder underlying lake (Figures 2(i) and (j); processes are detailed in Section 3.2). The temperature difference between the cold dense air above the lake and the warmer air over land drives the sea breeze. Indeed, with the warm air being more buoyant, it creates a low pressure zone that draws in cold air from over the lake onto the land. Without radiation, in S-A0, the sea breeze front continuously extends over land, and a stabilized state (i.e., when variables do not substantially change in time) is reached above the lake in 1–2 tsols.

The situation is different in S-A. The diurnally varying insolation heats the land during the day (see more details in Section 3.2). This leads to a relatively large sensible heat flux exchange between the hot land and the colder air above, especially when cold marine air moves inland (Figures 2(i) and (j)). The sensible heating of the atmosphere leads to dry convective overturning and instability. Convection is clearly visible in the surface wind during daytime as seen in the dark bands of the horizontal and vertical winds in Figures 2(a) and (e). Even though some convection is required over land to close the sea breeze circulation, the rigorous daytime turbulent episodes mix and diffuse the marine air mass over the land such that the sea breeze over the land largely collapses during the day (except at the shore). A new sea breeze forms at the lake shore on the following night and propagates inland until heating and mixing destroy most of the circulation on the next tsol. Because of the daytime attenuation of the sea breeze over land, the maximal sea breeze extension over land is limited to 700 km from the lake center (with a remnant, previous tsol front up to 1000 km). A similar turbulent convective phenomenon also occurs on Earth when cold sea breeze air masses propagate over heated surfaces (Crosman & Horel 2010).

When the diurnal variation of variables repeats itself every day, then the simulation has reached a stable, diurnally oscillating model state. For this reference simulation, this stable model state is reached after 1–2 tsols, as seen in Figure 2. Though diurnally varying, the intensity of the sea breeze wind is, on average, strongly increased in the case with radiation S-A (especially at night over land) compared to S-A0 without radiation (Figure 2). The addition of radiation increases the temperature difference between the air above the lake and above the land (explained in detail in Section 3.2), which leads to stronger winds toward the land, especially at night.

There is weak nocturnal turbulence over land in the case with radiation (Figures 2(a) and (e)). This turbulent convection is due to two processes. First, the marine air is substantially colder than the surface, even at night. Second, the land tends to be kept warm because the downward IR flux from the atmosphere is slightly higher than the upward IR flux emitted by the surface (Figure A1(b) in the Appendix). As is the case during the day, the temperature contrast between the surface and air above drives a sensible heat flux (Figures 2(i) and (j)) that destabilizes the lowest atmospheric layers (see more details in Section 3.2). This is different from what typically occurs on Earth, where the ground cools down very quickly by IR emission at night and other energy budget terms are unable to keep the land warmer than the air. Thus, the sign of the nocturnal sensible heat flux on Earth is usually opposite to that of Titan, which leads to a nocturnal radiation inversion on Earth and not on Titan. However, above the shallow unstable nocturnal layer on Titan, the marine layer is stable with a well-defined marine inversion, just like Earth.

### 3.2. Energy Budget

To investigate in greater detail the processes that drive the observed changes in structure, characteristics, and evolution between the radiative and nonradiative solutions, the evolution of key variables is compared to the evolution of energy fluxes in Figures 3 and 4. We identified wind, air temperature, land/lake temperature, and methane mixing ratio as key variables that affect the atmospheric circulation and energy transport, since these are variables that we initialize at the beginning of the simulation. The energy fluxes of sensible heat, latent heat, solar radiation, infrared radiation, and soil conduction then evolve as a response to both the initial state and the evolution of the key variables.

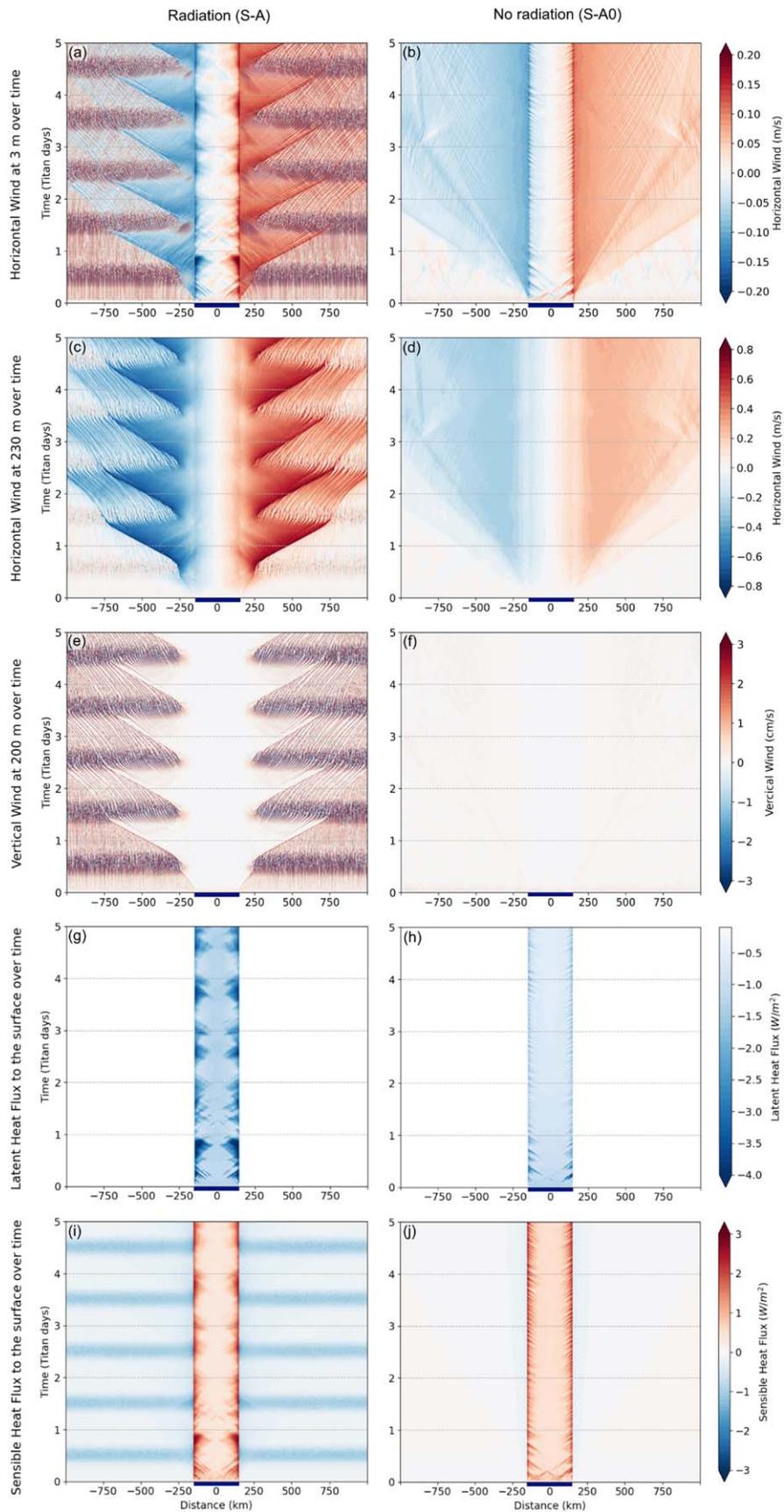

**Figure 2.** Evolution with time of the horizontal wind at 3 m (panels (a) and (b)) and 230 m (panels (c) and (d)) in altitude, the vertical wind at 200 m (panels (e) and (f)), the latent heat flux to the surface (panels (g) and (h)), and the sensible heat flux to the surface (panels (i) and (j)). Results obtained in the reference simulation with radiative transfer S-A (left column) are compared to results in the same conditions without radiative transfer S-A0 (right column). The dark blue line indicates the lake position. Simulations are started at midnight.

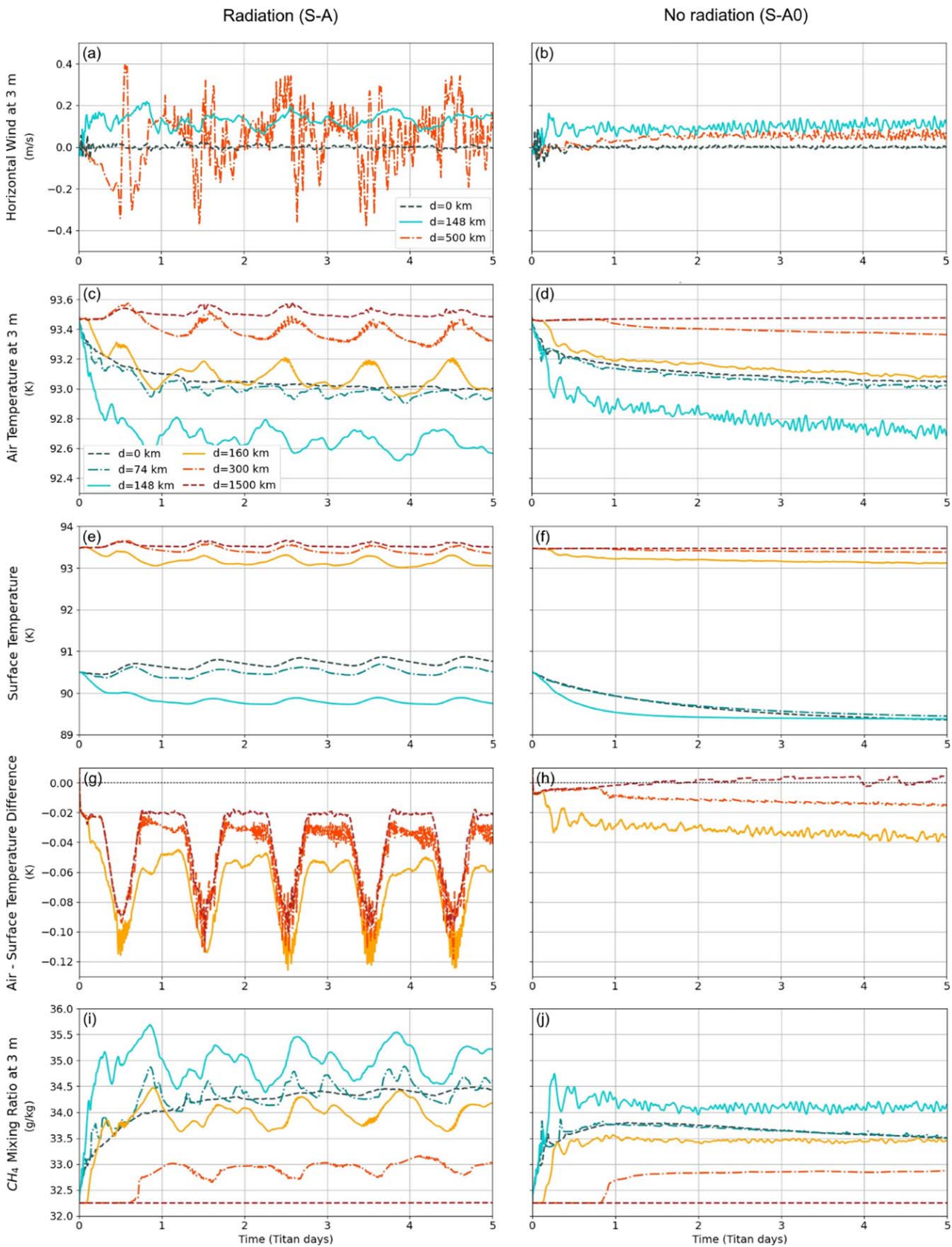

**Figure 3.** Evolution with time at the surface of wind (panels (a) and (b)), air temperature (panels (c) and (d)), land/lake temperature (panels (e) and (f)), air-surface temperature difference over the land (panels (g) and (h)), and methane mixing ratio (panels (i) and (j)) at different locations, with $d = 0$ km representing the center of the lake and $d = 160$ km the land just beyond the lake. Results obtained in the reference simulation with radiative transfer S-A (left column) are compared to results in the same conditions without radiative transfer S-A0 (right column). Not all lines are shown in panels (a), (b), (g), and (h) for readability. Simulations are started at midnight.

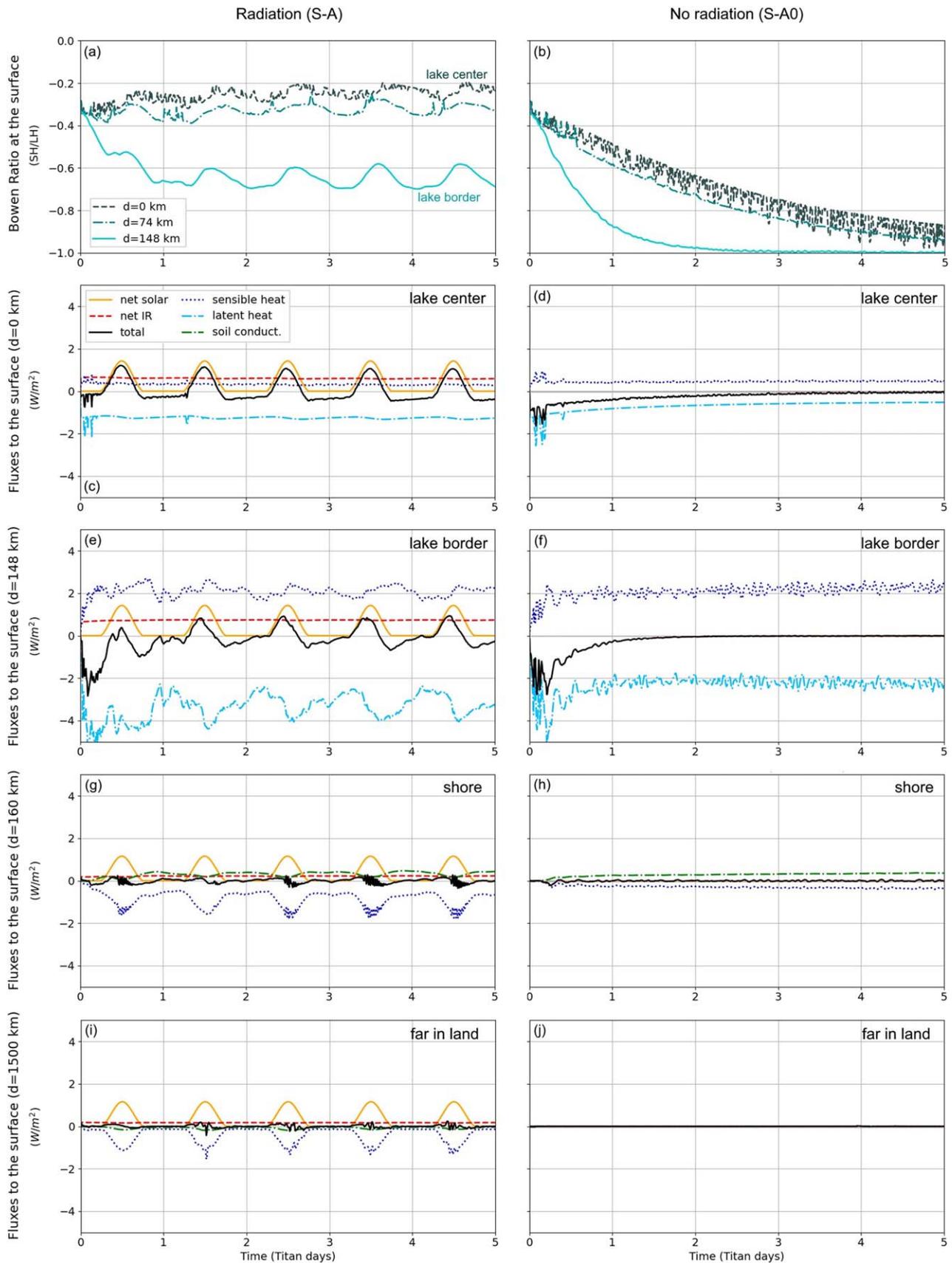

**Figure 4.** Evolution with time of the Bowen ratio (sensible heat flux SH over latent heat flux LH) over the lake (panels (a) and (b)), and the energy fluxes (solar and IR radiative fluxes, sensible and latent heat fluxes, and soil conduction heat flux) at different places: at the center (panels (c) and (d)) and border (panels (e) and (f)) of the lake, on the shore (panels (g) and (h)), and far inland (panels (i) and (j)). Results obtained in the reference simulation with radiative transfer S-A (left column) are compared to results in the same conditions without radiative transfer S-A0 (right column). Simulations are started at midnight.

### 3.2.1. Over the Lake

Over the lake, in the case without radiation, the latent and sensible heat fluxes are the only energy fluxes. Consequently, they are exactly opposite with a resulting Bowen ratio (sensible heat flux over latent heat flux) equal to −1, as previously predicted by Mitri et al. (2007) and modeled by Rafkin & Soto (2020). These prior works did not take solar and infrared radiation into account. Radiation adds new energy sources, and Figure 4 demonstrates that they should not be neglected. For instance, in the S-A simulation, the daily average solar radiation at the lake center is +0.4 $Wm^{-2}$, and the infrared radiation is +0.6 $Wm^{-2}$, while the sensible heat flux is +0.3 $Wm^{-2}$ and the latent heat flux −1.3 $Wm^{-2}$. At the lake border, the daily average solar radiation is also +0.4 $Wm^{-2}$, but the infrared radiation is +0.7 $Wm^{-2}$, the sensible heat flux is +2.1 $Wm^{-2}$, and the latent heat flux is −3.2 $Wm^{-2}$. At the stabilized state, the daily average of the surface temperature variation is null, and consequently the daily average of the sum of the energy fluxes to the surface is zero. As a result, the daily averages of heating (sensible heat, solar and infrared) and cooling (latent heat) fluxes are exactly opposite. The sensible heat flux changes very little with the addition of radiation at the lake border and slightly decreases at the center of the lake (from 0.48 to 0.31 $Wm^{-2}$). The conclusion from this behavior is that radiation has little direct impact on the sensible heat budget term over the lake. In contrast, the latent heat is found to be larger in magnitude with the addition of radiation: in daily average from −0.54 to −1.27 $Wm^{-2}$ at the lake center, and from −2.3 to −3.2 $Wm^{-2}$ at the lake border. As a consequence, the Bowen ratio varies between −0.7 and −0.2 (Figures 4(a)–(b)), which are values closer to what is commonly observed on Earth (see Roulet & Woo 1986, and the discussion in Rafkin & Soto 2020). The net energy flux at the surface of the lake is zero on average, but it evolves with the insolation. It is positive during the day, and negative during the night, heating the surface during day, and cooling it during night, consequently causing daily variations of the lake temperature (Figure 3(e)). We note that the IR flux has a very slight diurnal cycle due to the lake temperature variation (<0.02 $Wm^{-2}$, which is hardly visible in Figure 4).

The evolution of the different fluxes (Figure 4) can be linked to the variation of the variables observed in Figure 3. The solar and net infrared radiative flux heats the lake. A higher lake temperature in S-A compared to S-A0 drives the increase of the latent heat flux because of the dependence of saturation vapor pressure on temperature, and consequently a more rigorous methane evaporation process is found in the radiatively active simulation. Consequently, more methane vapor is observed above the lake in the simulation with radiation (compare Figures 3(i) and (j)). The moist air formed above the lake is also cooled by the sensible heat flux. Since the moist air above the lake is much colder than the air above the land, which is itself heated by sensible heat flux from the hotter land in the case with radiation (Figures 3(c) and (d)), stronger surface sea breeze winds (Figures 3(a) and (b)) form, which further enhances the evaporation process.

The intensity of energy exchange processes is not the same everywhere over the lake. Due to the symmetry of the system, the circulation has a near stagnation point over the center of the lake, and the fetch from the center of the lake to the shoreline allows the acceleration of the wind to reach a maximum value near the shore (Figures 3(a) and (b)). As a consequence of the stronger winds, the evaporation process is more efficient, leading to the coldest lake temperatures near the shore (Figure 3(e)). This in turn drives a strong sensible heat exchange at the border compared to the center of the lake (Figures 4(c) and (e)). The large evaporation near the shore plus the addition of methane from continuous evaporation on the air's trajectory from the center of lake produces a higher mixing ratio of methane and the coldest air at the lake border (Figure 3(i)).

### 3.2.2. Over the Land

On the land there is no evaporation (the latent heat flux is zero), but soil heat conduction results in an additional energy budget term not present over the lake. Heat conduction is highest at the shore. Indeed, the contact with the coldest marine air coming off the lake cools down the nearby surface land skin temperature (Figures 3(e) and (f)) by sensible heat transfer. The cold surface then drives a higher conductive heat flux from the warmer subsurface ground.

On the land, sensible heat flux with the colder overlying atmosphere tends to cool the surface, while the soil heat conduction and radiation act as heat sources (Figures 4(g)–(j)). In the case with radiation, the main energy transfer on land happens between the solar flux that heats the surface during the day, and the oppositely cooling sensible heat flux from the colder air above. During the night, the sensible heat flux with the atmosphere is smaller and acts in opposition to the soil heat conduction and the net infrared heating from the atmosphere. The observed air-surface temperature difference is always negative in the S-A simulation and is larger during the day than night (Figure 3(g)). The sensible heat flux from the Sun-heated land warms the lowest atmospheric layer (Figures 3(c) and (d)), and this leads to an enhancement of turbulence and convection during the day (Figure 2(e)). In the case without radiation, energy fluxes are effectively zero inland and the air-surface temperature difference stays close to zero (Figure 3(h)). Consequently, no turbulence is seen in that simulation. (Figure 2(f)).

### 3.3. A Reshaped Sea Breeze Circulation

Due to the diurnal radiative forcing, characteristics and evolution of the sea breeze circulation are modified between S-A0 and S-A. As seen in Figure 2, the stabilized phase is reached after 1–2 tsols. Figure 5 shows a vertical cross section of the wind circulation and the methane mixing ratio on tsol 4, which is representative of the stabilized, repeatable regime. For more details, a focus on the horizontal wind, vertical wind, and virtual potential temperature close to the lake is given in Figure 6.

Here again, a nonnegligible effect of radiative transfer is evidenced. Both the horizontal and vertical winds are more intense in S-A. Convective mixing is not present in S-A0, but it occurs both day and night in the radiatively active case (S-A), especially beyond the sea-breeze front. Within the marine layer, the convection is diminished, but still present. The strong daytime convective mixing is consistent with the prior discussion on the evolution of the energy budget terms. The overturning, convective circulations determine the height of the planetary boundary layer (PBL).

The transport of methane in the atmosphere is also different due to the different circulations. Methane is evaporated from the lake more efficiently near the lake shore in the radiatively active case where horizontal winds, the wind shear, and the horizontal temperature gradient are stronger. The methane is then transported over the land by the sea breeze. In S-A, the strong daytime vertical convection also more vigorously transports methane upward into the deeper PBL. The daytime turbulence is responsible for the higher methane mixing ratio above 400 m over land close to the shore than over the lake.

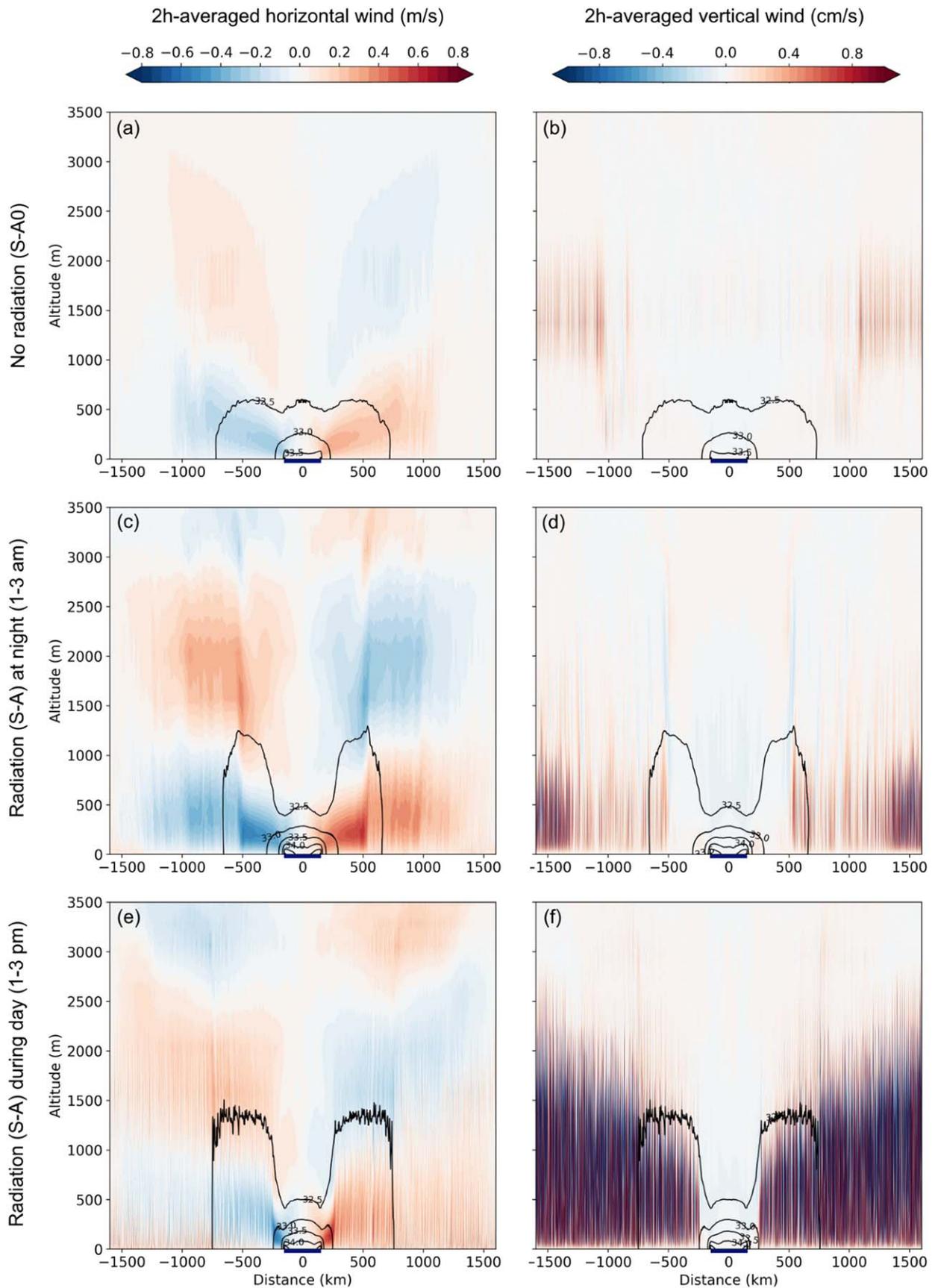

**Figure 5.** Horizontal and vertical wind along the 2D simulation. Contours give the methane mixing ratio every 0.5 g kg$^{-1}$. Average over 1–3 am (for S-A0 and S-A) and 1–3 pm (for S-A) on the fourth Titan day of simulation. The dark blue line indicates the lake position.

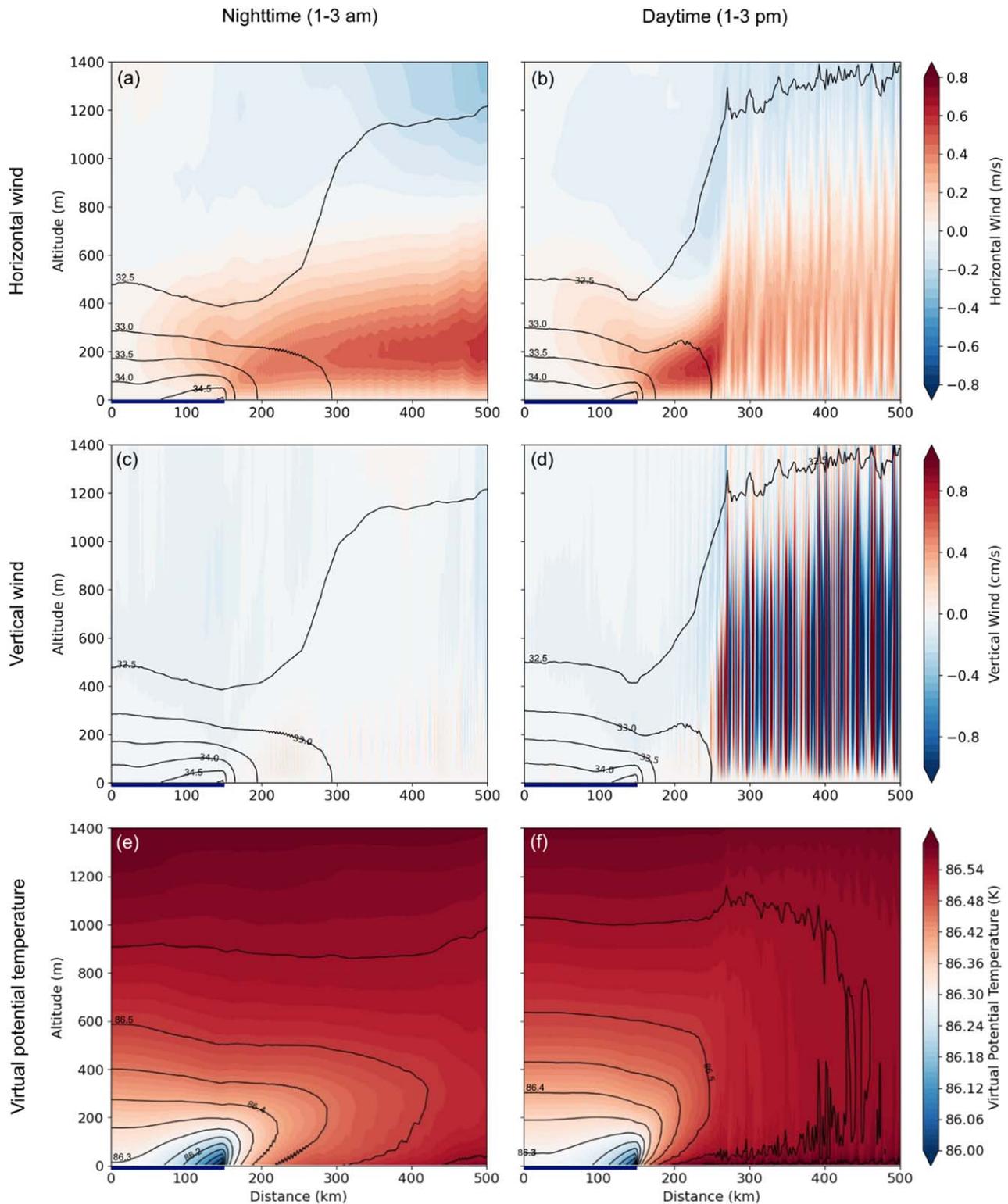

**Figure 6.** Zoom on the right lake section and nearby land. Horizontal wind (panels (a) and (b)), vertical wind (panels (c) and (d)), and virtual potential temperature (panels (e) and (f)). In panels (a)–(d), contours give the methane mixing ratio every 0.5 g kg$^{-1}$. In panels (e) and (f), contours give the virtual potential temperature every 0.05 K. The average over 1–3 am is shown in the left column, and that over 1–3 pm is shown in the right column, on the fourth day for simulation S-A. The dark blue line indicates the lake position.

The virtual potential temperature (Figures 6(e) and (f)) attests to the strong stability over the lake and the instability over the land (especially during the day). The case of the shore is more complex. Cold air is advected over the land, and the lowest layer is heated by sensible heat flux to create a shallow statically unstable layer (below 200–300 m during the night, and up to 400–600 m during the day) capped by a cold, stable, marine inversion. The stable marine layer hinders turbulence close to the shore while the depth of the shallow convective layer grows in depth as the marine air moves inland and warms.

In conclusion, the addition of the radiative transfer scheme in the reference case has a strong impact on the atmosphere-lake-land system. The new solar and infrared energy terms lead to a higher lake temperature, an enhanced latent heat flux, more methane evaporation, a departure of the Bowen ratio from −1.0, stronger winds, and a more efficient vertical mixing of the methane vapor with active convection and turbulence over the warmer land both day and night. In addition, all meteorological variables undergo diurnal variations.

## 4. Sensitivity Study

Most of the model input parameters have never been well constrained on Titan, including the lake mixed layer depth, the lake surface temperature and the underground temperature. In addition, some parameters should vary with time and location, such as the surface relative humidity and the background wind. Of course, insolation will also vary with location and season, but in a generally known way (see Section 5). In the previous section we detailed results from a reference case, where we selected given values for these parameters. In this section we evaluate the sensitivity of the model to these parameters, with a particular focus on the radiation-induced effects.

### 4.1. Underground Temperature: Control of the Surface Stability

In the reference case, the subsurface temperature is simply taken to be equal to the initial surface temperature (93.47 K), thereby instantiating a zero sensible heat flux initialization. In Rafkin & Soto (2020), the subsurface was not needed because there was no prognostic soil temperature model. The deep subsurface temperature should in theory be equal to the annual mean surface temperature, whereas the relevant subsurface temperature on the timescales appropriate for this study is the temperature at the depth of the diurnal thermal wave, which is usually shallower than the depth of the seasonal thermal wave. Huygens landed at 10.6°S just after the southern summer solstice, during the daytime. Therefore, one can expect the actual annual mean temperature to be less than 93.47 K. As discussed later in Section 5, diurnal and seasonal variations of the land surface temperature on Titan are very small, less than 0.3 K. We tested different values for the subsurface temperature (see Section 2) and chose to present here a reasonable "cold ground" case S-B characterized by a fixed underground temperature of 93.21 K, to be compared to a "warm ground" case of the reference simulation S-A. Any value of subsurface temperature between these two cases should be possible. Our simulations show that this 0.26 K difference between both simulations (leading to variations in the soil conducting flux of $\sim 0.25$ Wm$^{-2}$) has notable consequences on the resulting circulation, and thus, the model sensitivity to this parameter is large.

The circulation cross section of S-B (Figures 7(a)–(d)) may be compared to the previous S-A case (Figure 5). The daytime structure of the circulation is very similar in both cases, except that the top of the convective PBL (i.e., the top of the convection cells seen in the vertical wind) is at lower altitude in S-B. However, the nighttime circulation is notably different. In S-B, far inland at night, the net IR surface heating is compensated by the cooling soil conduction (Figures 7(e) and (f)). As a consequence, unlike in S-A, the land temperature stays lower than the air temperature, ensuring a sensible heat flux that drives stability, reducing vertical winds and eliminating turbulence over land at night. The absence of turbulence at night allows propagation of the sea breeze over the land with much less mixing than in S-A. The transport of methane in the atmosphere is also affected by this change in the circulation. Methane is advected deeper inland and is confined to lower altitudes.

Figure 8 shows the different behaviors of turbulence by plotting the Richardson number (Ri), defined as the ratio of the turbulent buoyancy source/sink to the wind-shear source. Negative values indicate a buoyancy source for convective PBL turbulence (i.e., statically unstable), while positive values indicate a buoyancy sink inhibiting turbulence. Negative Ri values are found from the surface to up to 1000 m over the land during daytime due to heating of the lowest level of the atmosphere through a sensible heat flux. During nighttime, S-A also has an unstable surface layer up to 800 m due to the IR heating of the surface and the subsequent heating of the lowest atmospheric layer by sensible heat flux (Figure 4(i)). Initially, S-B does not have an unstable nocturnal surface layer above the ground because of the cooling soil conduction process discussed above. However, the propagation of cold moist air from the lake over land leads to a shallow nocturnal unstable layer, visualized as a negative Ri. This lake-induced nocturnal unstable layer grows farther inland with each simulation day (compare Figures 8(a) and (c)). Nevertheless, it is capped by a stable cold marine inversion and consequently cannot trigger convection cells as during the day or as in S-A. Figure 8 also shows that the atmospheric stability inferred from virtual potential temperature positive gradients matches the stable positive Ri regions.

In conclusion, the underground temperature plays a major role in the stability/instability of the near-surface air during the night over land. A colder subsurface temperature leads to a colder land surface, often colder than the air, and higher atmospheric stability compared to the reference case. The intermediate cases between S-A and S-B are at least reasonable on Titan, and the real solution is difficult to determine without knowledge of the subsurface temperature. Nighttime turbulence affects the sea breeze structure and the mixing of methane within the atmosphere. However, in both cases, S-A and S-B, the qualitative evolution of most parameters (temperatures, winds, and Bowen ratio) is not significantly altered.

### 4.2. Initial Surface Relative Humidity: Control of the Evaporation Efficiency

The reference simulation was performed with a 45% initial surface relative humidity, which was the value measured near the equator by Huygens, corresponding to a methane mole fraction of 5.3% (Niemann et al. 2005). Evaporation variations with insolation combined with seasonally varying Hadley cells and the polar vortex strongly affect the polar moisture budget (Newman et al. 2016; Lora & Ádámkovics 2017), which indicates that variation of the surface humidity from the measured value is expected at different seasons and more distant locations. To gain insight on the effect of the initial humidity, three additional cases were tested: 0% (simulation S-C), 20% (simulation S-E), and 70% (simulation S-F) relative humidity (respectively, 0%, 2.4%, and 8.1% methane mole fractions). The 0% experiment can be compared with similar simulations in Rafkin & Soto (2020).

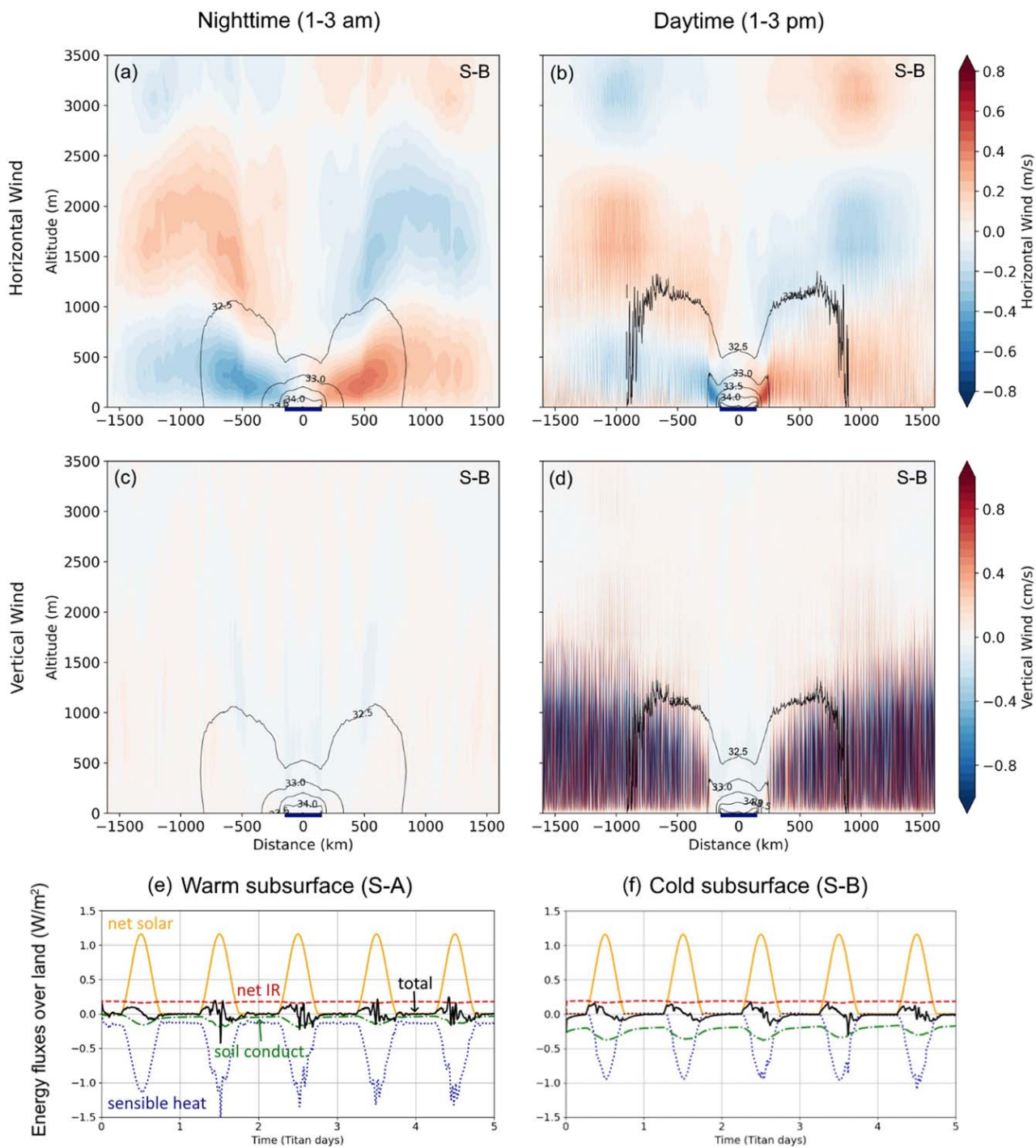

**Figure 7.** Horizontal (panels (a) and (b)) and vertical (panels (c) and (d)) wind along the 2D simulation S-B (cold subsurface). Contours give the methane mixing ratio every 0.5 g kg$^{-1}$. Average over 1–3 am (panels (a) and (c)) and 1–3 pm (panels (b) and (d)) on the fourth tsol. The dark blue line indicates the lake position. Panels (e) and (f) show energy fluxes over land far from the lake for simulations S-A and S-B.

The changes induced by an increase of methane surface relative humidity are detailed in Figure 9. At the highest ambient humidity (70%), the evaporation process is much less efficient. The latent heat flux over the lake, proportional to the difference between the saturation methane mixing ratio and the actual methane mixing ratio, decreases (two top lines in Figure 9). The direct consequences of the lower latent heat flux are a diminished production of methane vapor and a higher lake temperature (lines 4 and 6 in Figure 9). Since the lake is warmer, the sensible heat flux is lower (two top lines in Figure 9), and the air does not cool as much (fifth line in Figure 9). A second effect of this is the decrease of the IR net flux to the lake because the warmer lake in more humid conditions is radiating more upward, while the downward IR flux from the atmosphere does not significantly vary (two top lines in Figure 9). In conclusion, when ambient humidity increases, all energy fluxes become smaller except the solar flux, which thus has an increasing importance in the total energy budget. Consequently, larger diurnal variations in all model parameters are observed when compared with low-humidity cases. This effect is more important closer to the shore. For example, the surface wind and methane mixing ratio strongly follows a diurnal variation over the lake (see Figure 10).

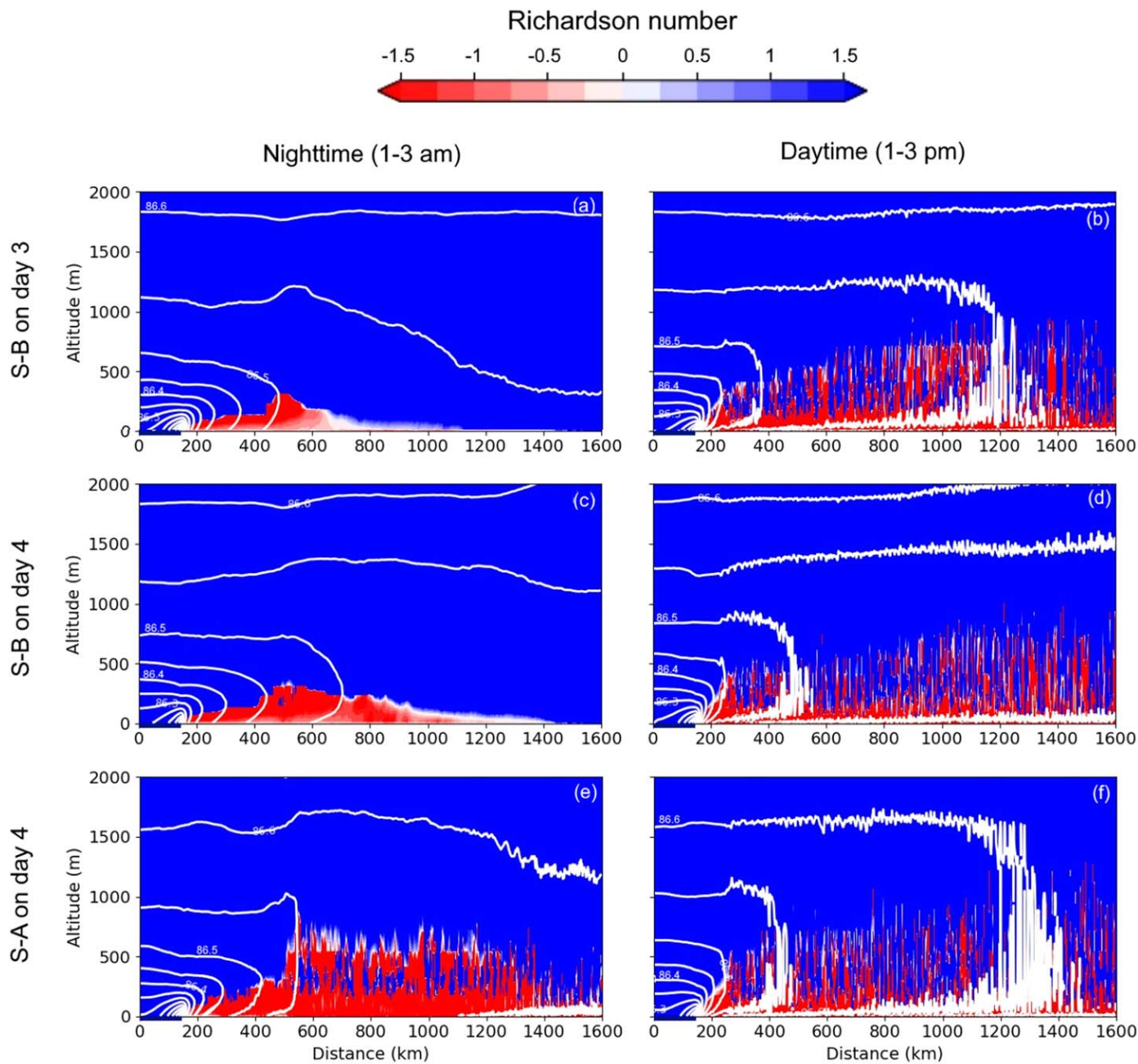

**Figure 8.** Richardson number along the 2D simulation. Contours give the virtual potential temperature every 0.05 K. Average over 1–3 am (left column) and 1–3 pm (right column) on the third (panels (a) and (b)) and fourth (panels (c) and (d)) Titan days of simulation for S-B, and on the fourth tsol (which is very similar to the third tsol) for S-A (panels (e) and (f)). The dark blue line indicates the lake position.

Proportionally, the decrease of the sensible heat flux is greater than the decrease in latent heat flux, leading to a Bowen ratio closer to 0 (third panel in Figure 9). This can be understood from an energetic point of view. For the system being stabilized, the diurnally averaged sum of the energy fluxes is zero. The sum of the solar and IR radiative flux and the sensible heat flux is equal in magnitude to the latent heat flux. The IR and sensible heat fluxes decrease in higher-humidity conditions, but not the solar flux. As a consequence, the latent heat flux decreases, but not as much as the IR and sensible heat fluxes.

The lake being less cold in high-humidity cases results in lower sensible heat exchange once the air moves over land, and the temperatures over the land close to the shore are warmer than in the reference case. In lower-humidity cases (0% and 20%), the daytime solar heating and the slight nighttime IR heating are not sufficient to trigger surface turbulence over the cold land shore. This is ever less so the case with increasing humidity, which has more limited cooling of the land and greater sensible heat warming of the lowest air layers, which allows for more convection. Convective cells consequently start closer to the lake as the humidity increases, as evidenced in the vertical wind (Figure 11). The presence of turbulence closer to the lake affects the sea breeze structure. As soon as it moves over land, the sea breeze front and the marine layer are mixed by the vertical winds. The consequence of the warmer marine layer and the mixing is that the sea breeze extension over the land is more limited in high-humidity cases (Figure 10).

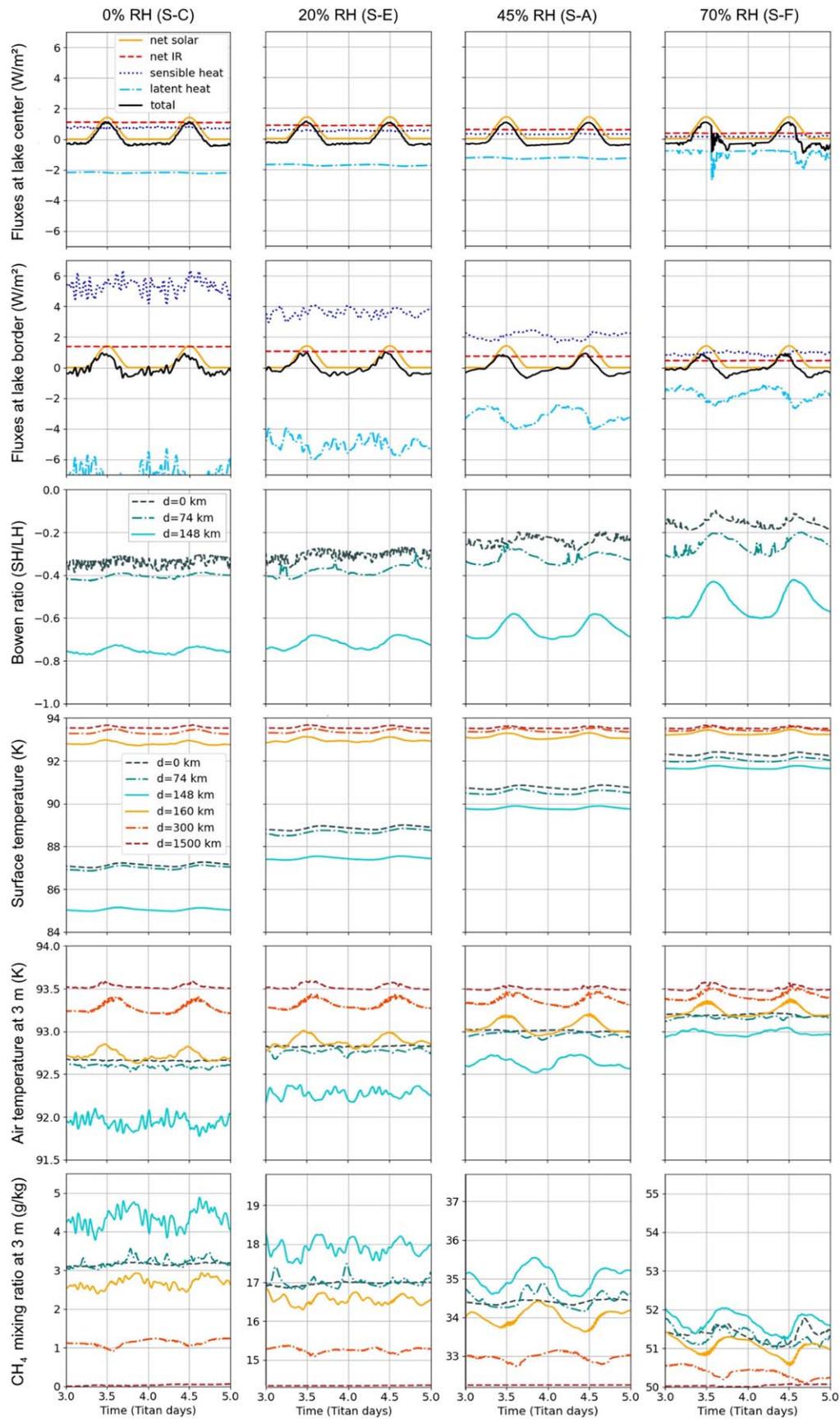

**Figure 9.** Evolution with time of the energy fluxes (solar and IR radiative fluxes, sensible and latent heat fluxes) at the center and border of the lake, of the Bowen ratio, the surface temperature, the air temperature at 3 m, and the $CH_4$ mixing ratio at different locations. Results shown are obtained in simulations S-C (0% relative humidity RH), S-E (20%), S-A (45%), and S-F (70%).

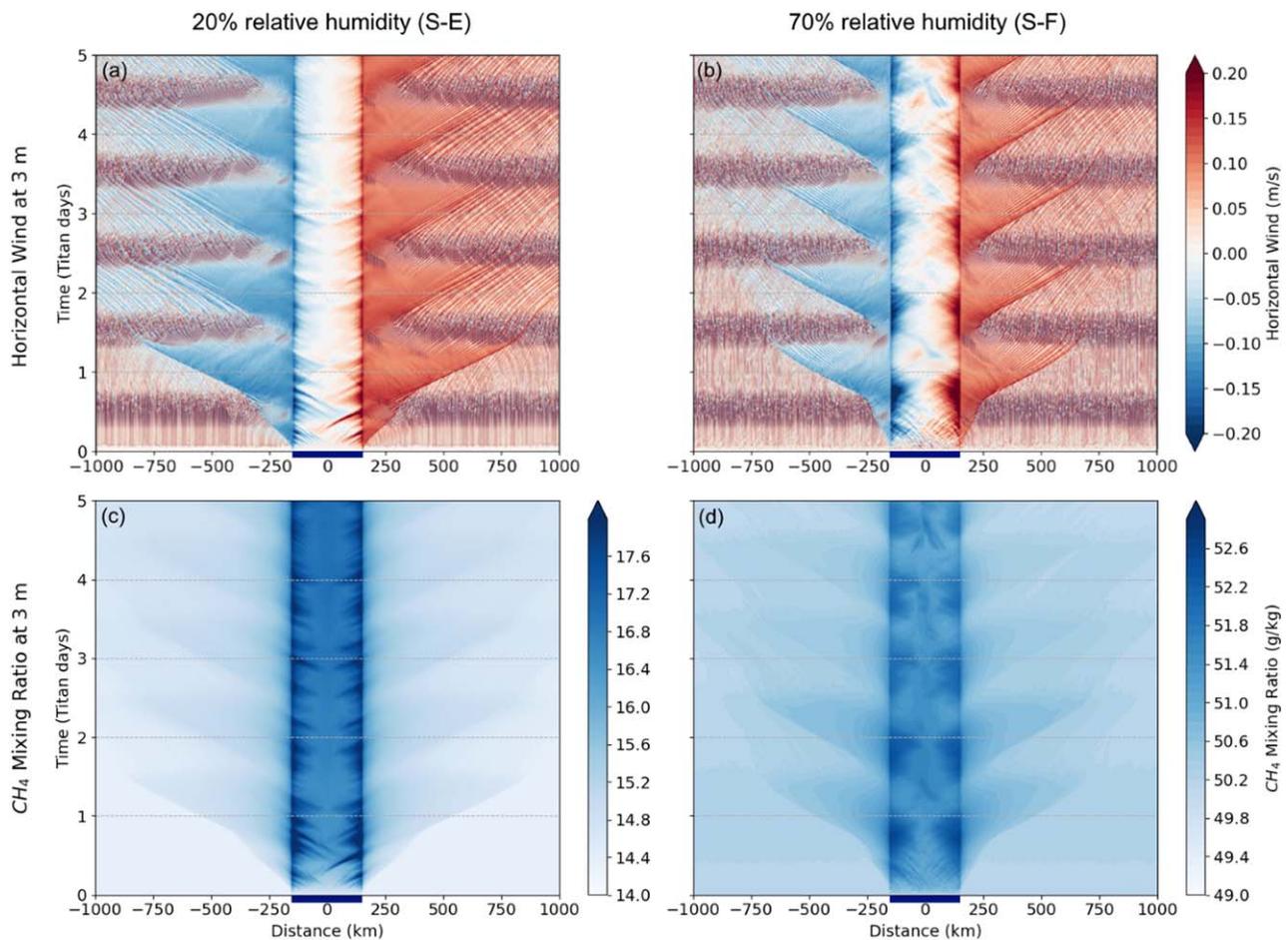

**Figure 10.** Evolution with time of the horizontal wind and the methane mixing ratio at 3 m altitude. Results shown are obtained with the simulations S-E (20% humidity) and S-F (70% humidity). The dark blue line indicates the lake position. Simulations are started at midnight.

Finally, the 70% humidity case shows an additional specific variation. On the initial methane profile, the altitude of the inflection point of a constant methane mixing ratio near the surface to a decreasing mixing ratio happens at a lower altitude than for the lower-humidity cases (Figure 1). The inflection is found between 3000 and 4000 m, which are altitudes reached by the sea breeze circulation. As a consequence, dry air is entrained by the circulation and descends over the lake (Figure 11). This phenomenon induces a slight general drying of the lower atmosphere with time (Figure 9 bottom-right plot). The Huygens/GCMS data in Figure 1 suggest that this could happen on Titan at lower relative humidity, but this does not happen with the idealized methane profiles.

In conclusion, the addition of initial methane vapor in the model strongly impacts the results. All energy fluxes, except the solar heating, are strongly decreased, leading to processes more strongly forced by the diurnal solar cycle. The main consequences at the surface are a warmer lake and less methane evaporation. The structure of the sea breeze circulation is also slightly modified with a more limited sea breeze extension over the land due to an enhanced turbulence close to the lake in the higher-humidity cases. Dry air coming down over the lake is possible in high-humidity cases.

### 4.3. Effects of Initial Lake Temperature, Depth of the Lake Mixed Layer, and Background Wind

The effects of the initial lake temperature, the depth of the lake mixed layer, and the background wind have already been investigated in Rafkin & Soto (2020). As shown in Table 1, we performed various simulations with the updated version of their model (with radiative transfer and soil conduction). As the general conclusions of these sensibility studies do not change, we mention here only the main changes induced by the additional physics. A reader in search of more details can refer to our Supplementary Information document and to all of our model outputs that are made available for additional analysis (see the Supporting Data section).

#### 4.3.1. Initial Lake Temperature: No Influence on the Stabilized State

Lake temperatures on Titan are not well constrained. However, Rafkin & Soto (2020) showed that lakes are likely to be much colder than the land due to long-term evaporative cooling, and that they could possibly be below their freezing temperature, although the freezing temperature is dependent on poorly constrained composition and lake dynamics. Yet, radar measurements of dielectric properties and brightness temperature indicate that most lakes are likely liquid (Stofan et al. 2007; Wye et al. 2009). We tested different values for the initial lake temperature (93.47 K, 90.5 K, and 88.0 K), and we found that the stabilized regime after 2–3 days is nearly identical in all cases (Figure 12). Only the initialization phases are different, but these do not vary with the addition of radiation.

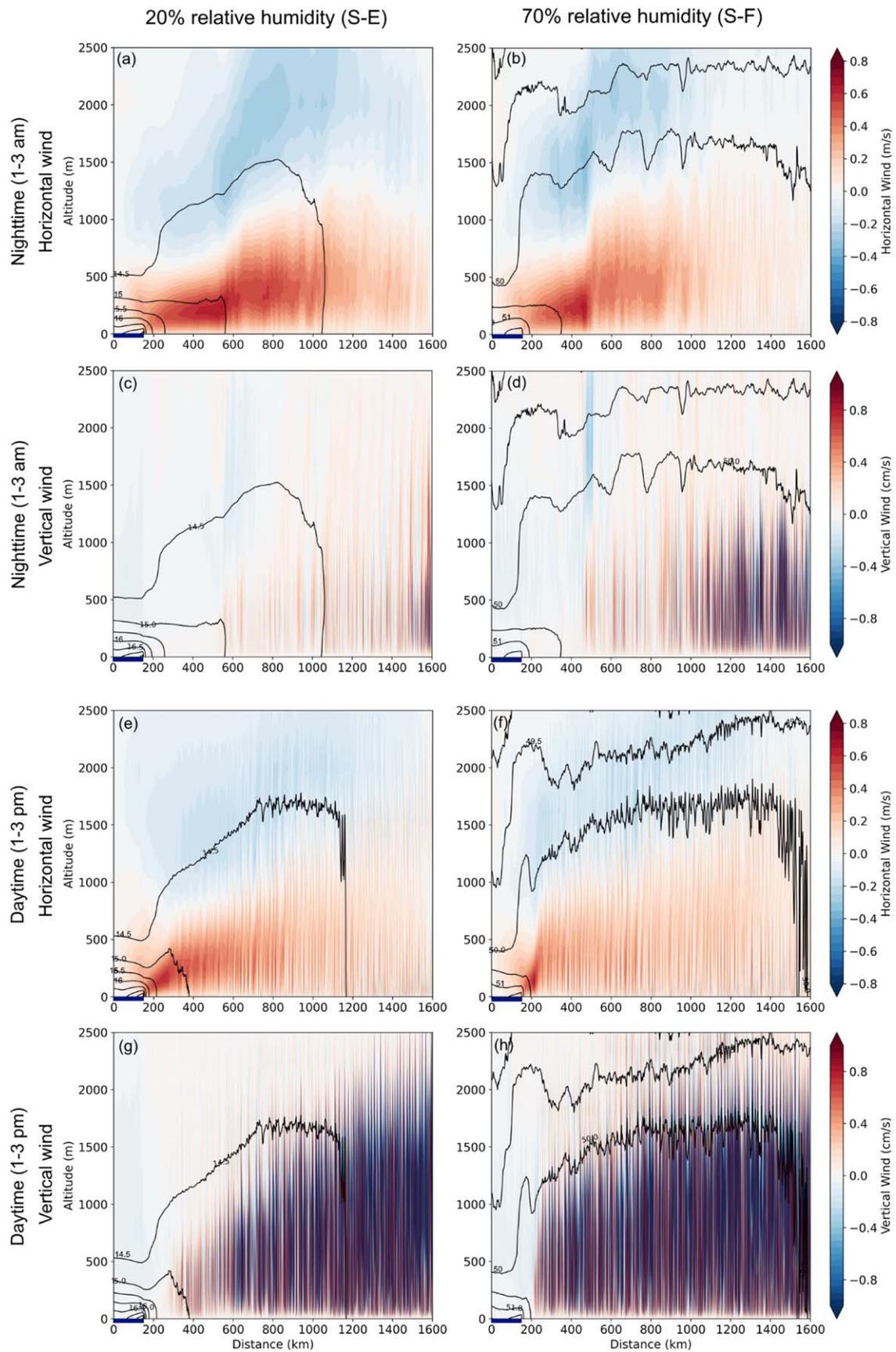

**Figure 11.** Horizontal and vertical wind along the 2D simulation. Contours give the methane mixing ratio every 0.5 g kg$^{-1}$. Averages over 1–3 am and 1–3 pm on the fourth day for simulations S-E and S-F are given. The dark blue line indicates the lake position.

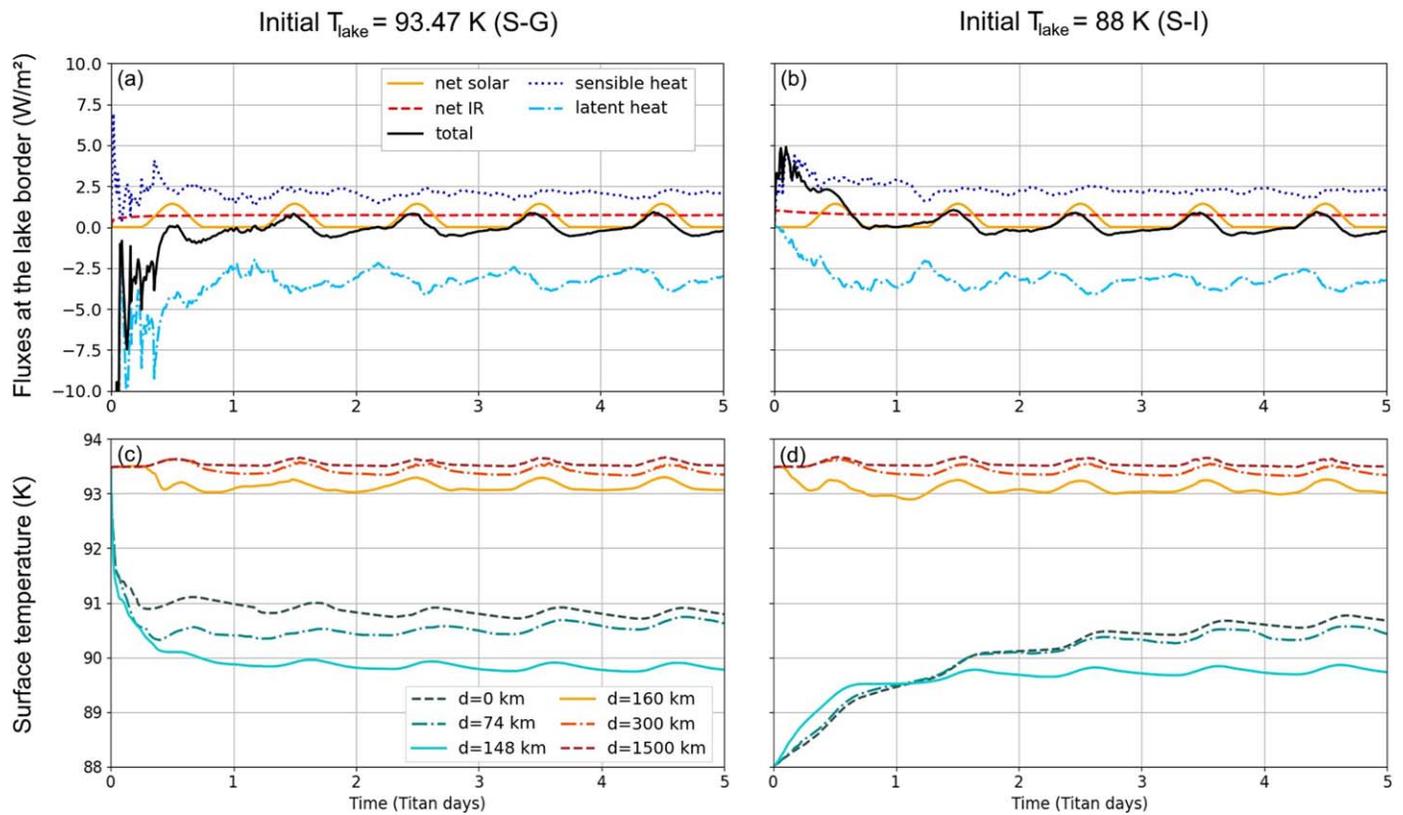

**Figure 12.** Output results from S-G ($T_{lake,ini}$ = 93.47 K) and S-I (88 K). Panels (a) and (b) show the time evolution of energy fluxes at the shore (to compare to Figure 4(e)). Panels (c) and (d) show the time evolution of surface temperature (to compare to Figure 3(e)).

The lake equilibrium temperature does not depend on the initial lake temperature. The analysis of all of the performed simulations shows that the initial relative humidity and the insolation (and the background wind to a lower extend) dictate the lake equilibrium temperature, which can then be predicted. If a meteorological or geologic phenomenon changes the lake temperature from its equilibrium value on Titan, it would tend to go back to it through surface flux forcing, especially through modulation of the methane evaporation rate. Only the initialization phases are different, but these do not vary with the addition of radiation.

*4.3.2. Depth of the Lake Mixed Layer: Control of the Inertia to (Diurnal and Seasonal) Variations*

Previous sections have focused on a shallow mixed layer of 1 m. However, lakes on Titan may be over 100 m deep (Mastrogiuseppe et al. 2019), and therefore the mixed layer could be much deeper than 1 m. In particular, pure methane lakes are expected to be highly thermally stratified, with colder layers at the bottom (Tokano 2005a). Consequently, the nighttime and wintertime cooling of the top of the lake leads to an overturn in the lake stratification, and so to a deeper mixed layer. In contrast, the lake stratigraphy is likely to be more stable (i.e., with a shallow mixed layer) during the day and in the summer, when the top of the lake is heated by the Sun. However, this is not totally straightforward as a higher insolation also triggers a stronger cooling by evaporation, which tends to produce cold, sinking fluid. Additionally, nonpure methane lakes (potentially mixed with ethane and/or nitrogen) and lakes with a composition varying with depth (Steckloff et al. 2020) could have a different equilibrium stratigraphy to cold at the bottom and warm at the top, triggering different values for the mixed layer depth (Tan et al. 2015). A more complex lake model with multiple layers could partially address the issues above, but the composition and pycnal behavior as a function of composition and temperature would still need to be known or specified.

The primary conclusion related to mixed layer depth is in complete agreement with results in Rafkin & Soto (2020): a deeper mixed layer acts as a higher inertia to changes in the lake, while shallower mixed layers and shallow lakes react more strongly to the effects of the daily cycle. This impacts mostly the lake temperature variations. Another consequence is the higher capacity of the lake to inject methane in the atmosphere for a given environmental change (e.g., seasons), because of their longer equilibrium time constant. On the other hand, we observe that the sea breeze structure stays exactly the same independently on the depth of the lake mixed layer.

*4.3.3. Background Winds: Asymmetry of the Sea Breeze*

We ran two simulations with background winds of 1 m s$^{-1}$ (S-L) and 3 m s$^{-1}$ (S-M). Although near-surface wind values of 1 and 3 m s$^{-1}$ are high for the sluggish atmosphere of Titan, these are the initial background winds at all altitudes in the model domain, and the surface friction quickly reduces the winds near the surface. The use of background winds quickly advects the lake-induced wind structures and methane vapor toward the downward edge of the simulation domain. For this reason, the simulation domain width was increased to 6400 km (with the 300 km lake at the center), and then integration time was reduced to 4 tsols. Periodic boundary conditions were retained, and no cyclic contamination was found in this configuration.

The addition of a background wind deforms the sea breeze structure, quickly diffuses cold air over land in the downwind direction, and creates an updraft front near the upwind shore. Surface winds also increased, leading to more methane evaporation and an accentuated cooling of the lake. In the 3 m s$^{-1}$ case, the sea breeze is nearly overwhelmed by the background wind. Nevertheless, all parameters and fluxes still show a distinctive diurnal variation (although attenuated).

## 5. Effect of Seasonal and Latitudinal Insolation Variations

The solar daily maximum insolation and the time of solar illumination depend on the latitude and solar longitude (season). To investigate dependencies, we investigated three extrema: the polar night (Ls 90°, latitude −85°, S-N); the maximum insolation point (Ls 270°, latitude −42°, S-O; Lora et al. 2011); and the polar day (Ls 270°, latitude −85°, S-P); as well as two equinox simulations at typical lakes latitudes: 70° (S-U) and 85° (S-V). These results are compared to the reference simulation S-A (Ls 0°, latitude 42°). These studies are only possible with the inclusion of radiative forcing, which was not considered in prior works. As discussed in Section 2, these are idealized simulations done with the same thermal conditions (initial thermal profile and constant subsurface temperature). These simulations aim to understand the direct (short-term) effect of insolation change on the lake evaporation processes, and they are not meant to exactly reproduce the thermal environment at all seasons and latitudes.

The intensity and extension of the sea breeze changes in the different cases (Figures 13 and 14). The evolution of key parameters given in Figure 15 helps to explain these differences. The intensity and the vertical extension of the sea breeze is increased at higher insolation. Indeed, higher solar radiations lead to a stronger heating of the land, which favors turbulence through latent heat flux. Higher and stronger convection cells are thus observed with more insolation. These diffuse the sea breeze front and increase its vertical extension. At higher insolation, the temperature difference between the air above the lake and above the land is larger. This leads to a more intense sea breeze, as determined by the wind speeds and thermal contrasts between the air masses on either side of the sea breeze front. The nighttime turbulence over the land in the diurnally varying cases is lower than during the polar night. This difference is because the air-surface temperature gap is larger during the polar night than during a diurnally varying simulation at night. The methane distribution in the atmosphere follows the sea breeze structures, going higher in altitude in higher-insolation cases, and spreading quickly over land at night.

Most parameters follow the insolation trend: a higher-insolation daily average leads to higher daily averages of latent heat flux, sensible heat flux, surface and air temperatures, and methane mixing ratio. These parameters also follow the diurnal variations of the solar flux, and show stronger diurnal variations in the cases with stronger insolation diurnal variations. Some have variations out of phase with the solar variations. In particular, the land temperature diurnal increase follows the insolation, but the lake temperature increase is shifted toward the afternoon due to the higher inertia required to heat liquid methane. The air temperature over the lake nearest the shore is also out of phase. It decreases in the afternoon due to strong sensible heat cooling of the evaporatively cooled lake.

In conclusion, seasonal and latitudinal variations in the solar flux affect the sea breeze circulation and have repercussions on the surface and air temperatures, and, therefore, on the methane budget. Comparisons between the two most extreme cases (maximum insolation S-O and polar night S-N) show that differences are not large, but still relatively important for the sluggish atmosphere of Titan: +1 K maximum at the lake center, +0.3 K on the land, +0.3 K maximum in the atmosphere at 3 m, +0.5 m s$^{-1}$ on the maximum horizontal wind at 230 m in altitude (thus doubling values in polar night), and +4 cm s$^{-1}$ on the maximum vertical wind (more than doubling the corresponding values in polar night).

Cassini measurements with the IR spectrometer CIRS provided insight into Titan surface temperatures with latitude and seasons (Jennings et al. 2011). An average seasonal temperature change of 0.5 K was observed between winter and spring, an order of magnitude coherent with the outputs of our simplified model. Regarding latitude dependence, these Cassini measurements found that the polar regions were ∼3–4 K colder (in average over all terrains) than the equator at 93.4 K. Our results show that the presence of many lakes at the poles could be part of the explanation, as lakes and close surroundings of lakes (i.e., the lake district) can be several kelvins colder than dry land due to evaporation.

## 6. Discussions

### 6.1. Evaporation and Insolation

Previous works (Aharonson et al. 2009; Lora et al. 2014; Newman et al. 2016) showed that the latitudinal distribution of lakes on Titan is closely linked to the annual insolation, with lakes concentrated at high latitudes, and especially at the pole with the milder summer. In agreement with many of these studies, we show here that this is probably not (only) due to a direct effect of insolation on the evaporation of lakes.

The level change of lakes due to methane evaporation ($\Delta h$) can be deduced from the latent heat flux (LH), the latent heat of vaporization ($l_v = 5.1\,10^5$ J kg$^{-1}$), and the liquid methane density at Titan temperatures ($\rho = 447$ kg m$^{-3}$) with: $\Delta h = \mathrm{LH}/(\rho \cdot l_v)$. Estimations issued from simulations done at different seasons and latitudes for shallow lakes (1 m mixing layer) and with 45% initial relative humidity are given in Table 2. $\Delta h$ is given in meters per Earth year, which is the unit used in the literature. Note that one Titan year is equivalent to 29.5 Earth years. We also note that in a 2D simulation, the proportion of shores on the lake area is underestimated. As winds and therefore latent heat fluxes are higher on the shores, the average latent heat flux over the lake in two dimensions may be underestimated compared to the reality in three dimensions from a geometric perspective. To give an upper limit, values at the shore are also indicated in Table 2.

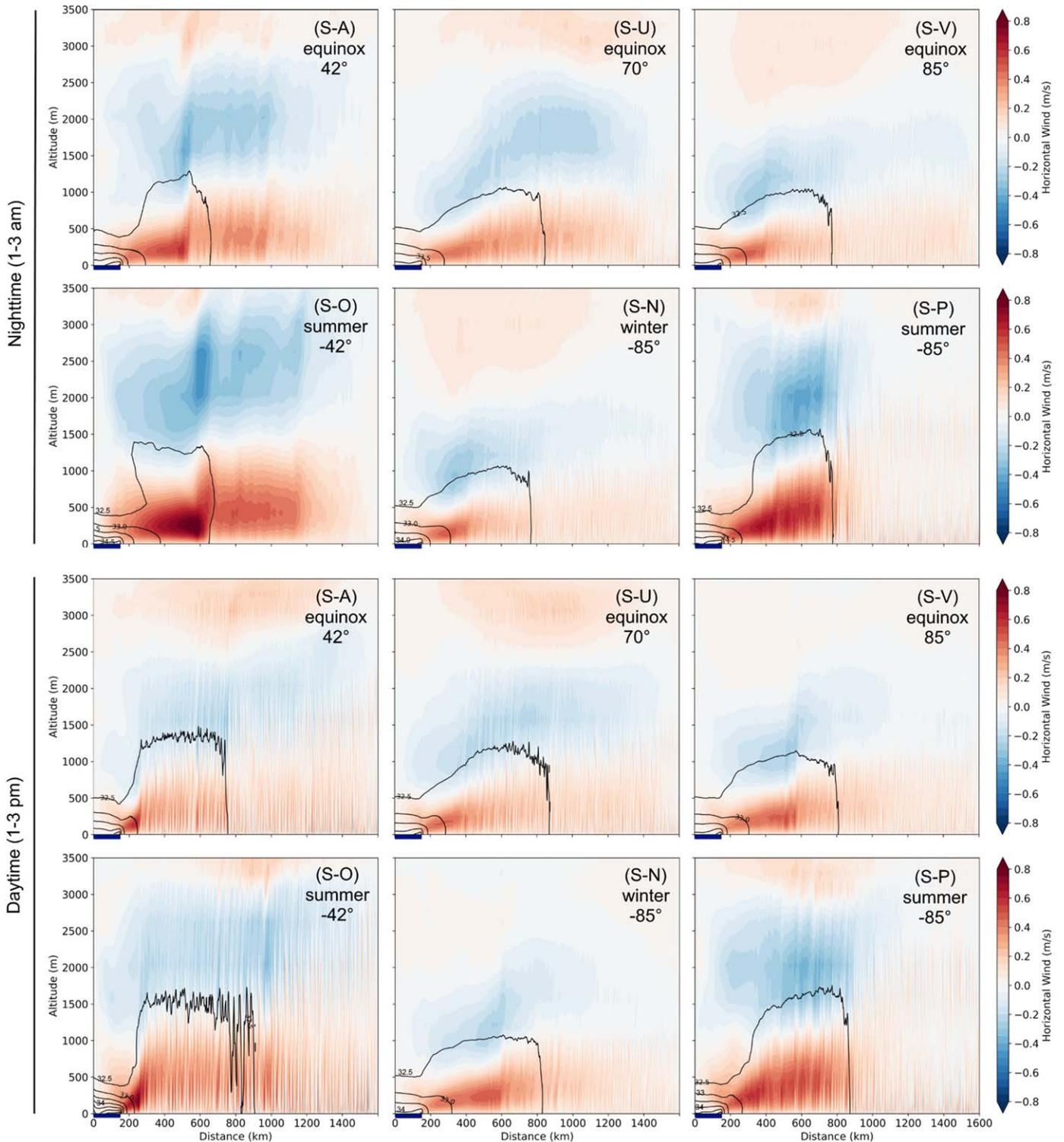

**Figure 13.** Horizontal wind along the 2D simulation. Contours give the methane mixing ratio every 0.5 g kg$^{-1}$. The averages over 1–3 am and 1–3 pm on the fourth day for simulations S-A, S-U, S-V, S-O, S-N, and S-P are given. The dark blue line indicates the lake position.

The evaporation values are in the low range of what was expected by Mitri et al. (2007) using overestimated fluxes (0.3–10 m yr$^{-1}$). We observe that S-A and S-R give nearly the same evaporation with a different daily averaged insolation. This means that not only is the average insolation important, but so is the maximum insolation. The evaporation efficiency is thus not linear with the insolation. The maximum insolation reached in one complete year is always smaller at the poles than at lower latitudes. This, among other things, tends to dry the lower-latitude regions.

Due to the eccentricity of Titan's orbit around the Sun, the south pole undergoes a higher evaporation rate in the summer than the north pole. However, the northern summer lasts longer than the southern summer. Schneider et al. (2012) investigated this effect with a general circulation model (GCM) and obtained a decreased evaporation over a Titan year between two poles of 0.10 m. This value is consistent in order of magnitude with our observations (though we cannot directly compare because we did not integrate over one Titan year).

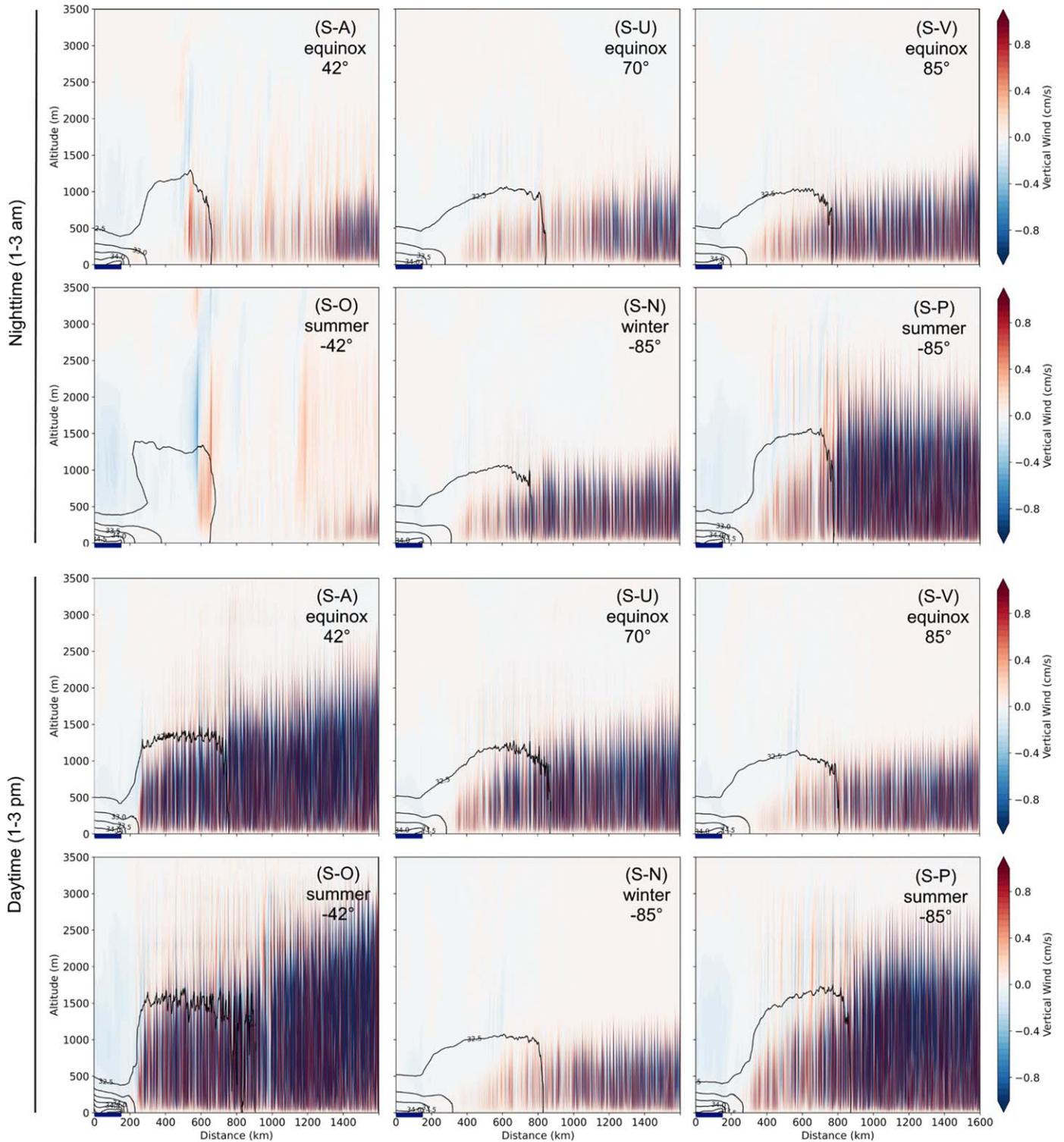

**Figure 14.** Vertical wind along the 2D simulation. Contours give the methane mixing ratio every 0.5 g kg$^{-1}$. The averages over 1–3 am and 1–3 pm on the fourth day for simulations S-A, S-U, S-V, S-O, S-N, and S-P are given. The dark blue line indicates the lake position.

The results given in Table 2 are in the case of a shallow mixed layer (1 m) and 45% initial relative humidity. However, we saw in the previous sections that evaporation is enhanced

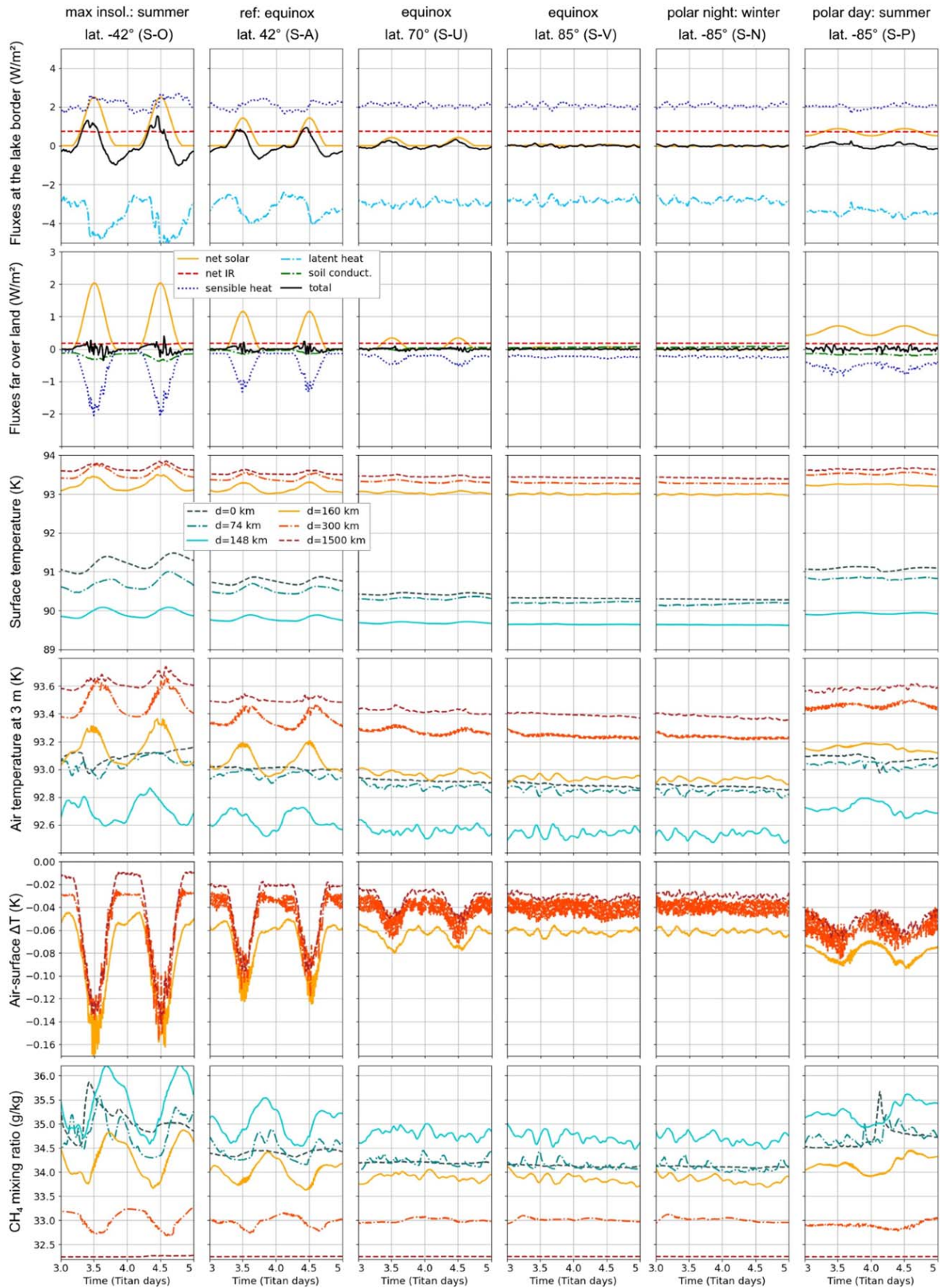

**Figure 15.** Evolution with time of the energy fluxes (solar and IR radiative fluxes, sensible and latent heat fluxes) at the border of the lake and over the land, the surface temperature, the air temperature at 3 m and the CH$_4$ mixing ratio at different locations. Results shown are obtained in simulations S-O (maximum insolation), S-A (reference, equinox at latitude 42°), S-U (equinox at 70°), S-V (equinox at 85°), S-N (polar night), and S-P (polar day).

Table 2
Comparison of Evaporation and Insolation Outputs at Different Seasons and Latitudes.

|  | S-N (Polar Night) | S-A (Equinox at lat 42°) | S-S (Winter at lat −42°) | S-O (Summer at lat −42°) | S-R (Summer at lat 85°) | S-P (Summer at lat −85°) | S-T (Ontario Lacus) |
|---|---|---|---|---|---|---|---|
| LH (W.m$^{-2}$): averaged over the lake/on the shore | −1.4 /−2.8 | −1.8 /−3.2 | −1.5 /−2.9 | −2.3 /−3.7 | −1.8 /−3.3 | −2.0 /−3.4 | −3.5 /−6.1 |
| $\Delta h$ (m yr$^{-1}$): averaged over the lake/on the shore | −0.20 /−0.39 | −0.25 /−0.44 | −0.21 /−0.40 | −0.31 /−0.51 | −0.25 /−0.46 | −0.28 /−0.47 | −0.49 /−0.84 |
| Daily averaged insolation (W.m$^{-2}$) | 0 | 0.39 | 0.08 | 0.82 | 0.55 | 0.69 | 0.71 |
| Maximum insolation (W.m$^{-2}$) | 0 | 1.43 | 0.40 | 2.51 | 0.70 | 0.88 | 1.42 |

**Note.** Simulations are for shallow lakes (1 m mixing layer), with 45% initial relative humidity at the surface, except for S-T, which is for a 100 km large lake, a 30 m mixing layer, 20% initial humidity, and summer at 72°S. The latent heat flux LH and the lake level change $\Delta h$ are averaged over the lake and over the simulation days 4 and 5.

for deeper mixed layers, low-humidity cases, and cases with background wind. Compared to S-A, LH (and so $\Delta h$) is increased by 40% in the 100 m mixed layer case (S-K), by 40% in the 20% humidity case (S-E), and by 15% in the 1 m s$^{-1}$ background wind case (S-L). Therefore, a combination of deep mixed layer, low humidity, and background wind could realistically at least double the values given in Table 2.

Observations in the south pole in early southern summer and southern autumnal equinox suggest that the level change of some lakes could reach ∼1 m yr$^{-1}$ (Turtle et al. 2009; Hayes et al. 2011). This is higher than the evaporation rates found in this work in S-P (2D—45% humidity—1 m mixing layer). However, a combination of 3D geometrical effects, deeper mixing layer, low humidity, and background wind could realistically allow the evaporation rate to approach ∼1 m yr$^{-1}$. The S-T simulation combines most of these parameters (summer, 20% humidity, and 1 m s$^{-1}$ background wind, but still in two dimensions) in the case of Ontario Lacus (72°S, 100 km large, 30 m mixed layer). The evaporation rate nearly double compared to S-P, and could explain the major part of the observed evaporation rate. We emphasize that our simulations are idealized and that results should be viewed in this idealized context. For example, Ontario lacus is not a pure methane lake (see discussion in Section 6.3). Regardless, the main conclusion here is that a lower humidity and an increased background wind can notably increase the evaporation rate.

Other factors are also very likely to influence or even drive the accumulation of moisture at the North pole, like enhanced precipitations driven by the global circulation (Aharonson et al. 2009; Schneider et al. 2012; Lora et al. 2014; Lora & Mitchell 2015; Tokano 2019). However, cloud observations (Turtle et al. 2018) and recent models (Lora et al. 2022) do not completely support this, and more in detail investigation is needed to understand the effect of lake-induced circulation on deep convection and precipitation.

The determination of realistic lake evaporation rates is complex. While the levels of most large lakes do not appear to vary with seasons, there is proof of emptying lakes (Hayes et al. 2011; MacKenzie et al. 2019a). The current explanation is that small isolated lakes could be sensitive to insolation variations and seasonally dry out, while most lakes are suspected to be linked by a subsurface methane table, which consists in a huge methane reservoir capable of absorbing quick changes in lake levels (Hayes et al. 2008). The presence of the methane subsurface table is supported by cloud observations compared to the output of a Global Climate Model coupled to a methane table model (Lora & Ádámkovics 2017; Turtle et al. 2018; Faulk et al. 2020).

The presence of a methane subsurface table would suggest that lands close to lakes could be soaked with methane, and possibly also a source of methane evaporation. Similarly, wetlands found at lower latitudes than lakes could also be subject to evaporation. It is possible that evaporating, localized wetlands could drive a sea-breeze-like circulation. We plan to investigate this in a future work. The result is of particular importance for the coming Dragonfly mission, which will explore Titan's surface by 2034 with a dual-quadcopter (Lorenz et al. 2018; Barnes et al. 2021). It will fly once every Titan day to another exploration site a few kilometers away. No orbiter will accompany the rotorcraft lander, and Dragonfly will have to rely on climate models to prepare its operations. The dense 1.5 bar atmosphere on Titan implies that even small winds can be powerful and need to be taken into account during the Dragonfly mission. As shown in this paper, winds up to 1 m s$^{-1}$ can be caused in altitude by evaporative effects over lakes. This number is to be adjusted with future 3D simulations. This effect is likely to be much weaker over wetlands than lakes. Nevertheless, even winds 10% of this intensity will need to be taken into account in the preparation of Dragonfly's operations (Lorenz 2021).

### 6.2. Surface Wind

The only in situ data we have—at the equator—suggest near-surface winds inferior to 0.2 m s$^{-1}$ from Huygens cooling rate after landing (Lorenz 2006). The absence of waves at the surface of lakes in nearly all of the radar observations implies a surface wind speed inferior to the threshold speed of 0.4 m s$^{-1}$ (Hayes et al. 2013). In addition, surface winds from GCMs (Tokano 2009a; Tokano & Lorenz 2015) also suggest an average value around 0.2–0.3 m s$^{-1}$. Our simulations are in perfect agreement with these low-wind values, as we find average surface winds inferior to 0.2 m s$^{-1}$, with a maximum around 0.4 m s$^{-1}$.

Nonetheless, "high" wind cases are not inconceivable over Titan's lakes. First, Tokano (2013) demonstrated with a GCM that cyclones over polar seas are not impossible and could enhance winds in this region. Second, winds could be locally accelerated by topography, irregular shorelines, and the interference between sea breezes coming from different lakes in the highly populated lake district around the north pole. This cannot be investigated with our 2D model, but the future development of a 3D model, including the Coriolis force, topography, and the irregular shorelines of lakes is a necessary step to investigate the possibility of local high surface wind cases.

### 6.3. The Lake Composition

The current version of the model assumes pure methane lakes. But lakes on Titan are known to also include some ethane, lesser amounts of other volatiles, and dissolved nitrogen in varying proportions. In the north, lakes are mainly composed of methane (∼70%), with the remaining 30% equivalently shared between ethane and nitrogen (Mastrogiuseppe et al. 2016, 2019). However, Ontario Lacus in the south is more concentrated in ethane (∼50% methane, ∼40% ethane, and ∼10% nitrogen; Brown et al. 2008; Mastrogiuseppe et al. 2018).

We showed that pure methane lakes are strongly subject to evaporation and induce a sea breeze due to the evaporative cooling. However, ethane-rich lakes are less subject to evaporation and cooling due to the lower saturation vapor pressure of the methane-ethane ideal mixture, as described by Raoult's law (see discussions in Mitri et al. 2007 and Rafkin & Soto 2020). With daily and/or seasonal variations, the lakes could remain warmer than the surrounding lands at night. Then, the circulation would reverse to a land breeze with wind blowing from the cooler land toward the warmer lake (Tokano 2009a). This is typical of what happens on Earth, where water lakes and seas keep the heat from the day/summer and become heating sources at night/in winter, creating land breezes. The possibility of a reversed wind circulation over ethane-rich lakes is still to be investigated. However, the high-methane-humidity case may closely mimic this scenario to a certain extent, since high relative humidity naturally reduces methane evaporation and lake cooling (see discussion in Rafkin & Soto 2020). A land breeze was not found in the high-humidity simulation.

Pure methane and pure ethane have a freezing point near 91 K. Therefore, lakes should mostly freeze in our simulations. However, when methane and ethane are mixed, they can remain liquid down to 72 K (Hanley et al. 2017). The temperature values found in this work are consistent with a liquid as soon as there is a few percent ethane mixed with methane. In the purest lakes recently formed by methane rain, we show that the shores would be the most likely to freeze, because they reach colder temperatures. In pure methane lakes, methane ice sinks; therefore, such lakes could freeze from below (Tokano 2005a, 2009b). However, as soon as some ethane is mixed with methane and if the temperatures are below the freezing point of pure methane, the ice formed is likely to float. There are only very few observations suggesting the possibility of floating ice (Hofgartner & Lunine 2013). Consequently, our model, which does not allow the possibility of ice to form, is reasonably representative of the most common observations of liquid lakes on Titan.

### 6.4. Moisture, Clouds and Rain

The evaporation of Titan's methane lakes is in many regards different from the evaporation of water from Earth's lakes. First, on Titan, the solar insolation is much weaker in magnitude and has a longer daily and seasonal variation timescale. Because of a very different saturation pressure, the evaporation rate can be much larger with methane on Titan, and the atmosphere can hold much more methane on Titan than water on the Earth. The atmosphere thus accumulates methane moisture, while drying the surface. Then, when the critical humidity is reached, cumulus convection starts, forming clouds and heavy rains (Hueso & Sánchez-Lavega 2006; Barth & Rafkin 2007; Hayes et al. 2018). In conclusion, Titan undergoes droughts and heavy storms. Hayes et al. (2018) noted that Earth's future long-term climate could more resemble that of Titan today: a warmer atmosphere will be able to hold more moisture. This will allow longer droughts interspersed by heavier storms. Also, lakes on Titan represents one-seventh of the atmospheric methane. Lakes have therefore less influence on the climate than oceans and lakes on Earth (Hayes et al. 2018). With the methane underground reservoirs, there is possibly more methane available to the atmospheric methane cycle (see Section 6.1). However, we do not know yet how much is stocked, and to what extend part of it could evaporate.

Clouds and rain are not simulated yet in our simplified 2D model. However, we show that the relative humidity is only slightly locally enhanced due to the lake evaporation. Thus, these simulations do not generated conditions conducive to clouds generated by lake processes. Some observations seem to indicate clouds formed by nearby lakes (Brown et al. 2009a, 2009b; Turtle et al. 2009). Nevertheless, the general circulation and local topography could also play a role in the formation of these clouds. And the presence of the observed clouds near lakes could sometimes be a coincidence. For now, our work shows that the lake–cloud connections only based on evaporation and sea breeze dynamics are rather unlikely to generate clouds. However, we plan to implement cloud microphysics in future studies and explore fully 3D domains with the addition of topography. These additions could concentrate humidity and forcing in locations that might force clouds. Tokano (2009a) also suggested that ethane lakes, by the mean of land breezes, could be a convergence center for moisture, and hence clouds and rain in the warm season.

### 7. Conclusion

To conclude, solar insolation and infrared radiation cannot be neglected on Titan when dealing with lakes. Though notably small in magnitude compared to the Earth, radiation fluxes on Titan are of the same order of magnitude as the other energy fluxes involved in the evaporation of lakes and the local circulation (i.e., the latent and sensible heat fluxes).

With our simplified 2D mesoscale model around a pure methane lake on Titan, we showed that the implementation of radiation changes:

[1] the energy balance and Bowen ratio, with two new energy sources: solar and infrared radiative fluxes;

[2] the evaporation of methane, which is intensified;

[3] the lake temperature, which is strongly increased (e.g., up to +1.5 K in a typical 45% surface humidity case). We also noted that it is almost completely determined by the initial relative humidity and insolation conditions;

[4] the local atmosphere dynamics. With radiation, there is a stronger sea breeze with diurnally varying wind intensity, and the formation of turbulence over land, which is strong during the day due to solar surface heating, and faint during the night, due to infrared cooling of the surface; and

[5] the methane distribution in the atmosphere. Methane is efficiently evaporated nearest the lake shores, advected inland over great distances during the night, and vertically mixed through the PBL during the day.

The quantification of the lake evaporation rate, the wind speed, and the lake equilibrium temperature is important for the improved interpretation of Cassini's observations, a better understanding of Titan's methane cycle, and the preparation of new missions. Our mesoscale model above a Titan lake provides increasingly robust insights for the above applications.

A.C. has received funding for this project from the European Union's Horizon 2020 research and innovation program under the Marie Sklodowska-Curie grant agreement No. 101022760. The effort of coauthors S.R. and A. Soto was supported through their institutional overhead, through a substantial donation of their personal time, and specifically not through any government funding. R.H. was supported by grant PID2019-109467GBI00 funded by MCIN/AEI/10.13039/501100011033/ and by Grupos Gobierno Vasco IT1366-19.

## Data Availability

The Fortran source code of the radiative transfer module developed for this work, all of the netCDF simulation outputs, and the Python codes to plot figures from the netCDF output files are available on Zenodo at https://www.doi.org/10.5281/zenodo.7023670. The Supporting Information document on the performed sensitivity study is available at the same link.

## Appendix A
## Description of the Radiative Scheme

### A.1. The Solar Insolation at the Top of the Atmosphere

The solar insolation at the top of the atmosphere and the solar zenith angle are computed at each time step. They depend on the season, the latitude, and the local time.

The solar flux actually reaching the top of Titan's atmosphere depends on the solar longitude $L_s$ and the orbital parameters of Titan and Saturn.

$$S_{TOA}(L_s) = S_0 \left( \frac{1 + e \cdot \cos(L_s - L_{s0})}{1 - e^2} \right)^2 \quad (A1)$$

where $S_0 = 15.04$ W m$^{-2}$ is the solar constant at the distance $(R_{aphelion} + R_{perihelion})/2$, $e$ is the eccentricity of the orbit of Saturn, and $L_{s0}$ is the solar longitude at the perihelion.

The incident solar flux is inclined to the vertical. The related parameter is $\mu_0$, the cosine of the solar zenith angle:

$$\mu_0 = \sin(\text{lat})\sin(\delta) + \cos(\text{lat})\cos(\delta)\cos(h) \quad (A2)$$

where $\delta = \epsilon \cdot \sin(L_s)$ is the declination, $\epsilon$ is the obliquity of Titan to the Sun, and $h = \pi(1 - \text{local hr}/12)$ is the hour angle.

### A.2. Absorption, Scattering, and Surface Reflexion of the Solar Radiation

The solar flux undergoes absorption and scattering in the atmosphere and is partly reflected at the surface.

The absorption is implemented with the extinction optical depth:

$$\tau(z) = \tau_0 \left( \frac{P(z)}{P_0} \right)^\gamma \quad (A3)$$

where $\tau_0$ is the surface optical depth, $P_0$ is the surface pressure, and $\gamma = 0.21$ is an empirical exponent (see supplementary information of Schneider et al. 2012).

We consider the multiple anisotropic scattering of a collimated beam in a homogeneous plane parallel atmosphere (Liou 1980; Thomas & Stamnes 1999). Fluxes depend on the azimuthally averaged radiative field. The azimuthally averaged radiative transfer equation in our case is therefore:

$$\mu \frac{dI_d(\tau, \mu)}{d\tau} = I_d(\tau, \mu) - \frac{\omega_0}{2} \int_{-1}^{1} p(\mu', \mu) I_d(\tau, \mu') d\mu'$$
$$- \frac{\omega_0 S_{TOA}}{4\pi} p(-\mu_0, \mu) e^{-\tau/\mu_0} \quad (A4)$$

where $I_d$ is the diffuse intensity, $\omega_0$ is the asymmetry factor, and $p$ is the scattering phase function.

We solve this equation using the two-stream approximation, which leads to the resolution of two coupled equations on $I_d^+(\tau)$ and $I_d^-(\tau)$, the averages of $I_d$ over the upward and downward hemispheres:

$$\overline{\mu} \frac{dI_d^+}{d\tau} = I_d^+ - \omega_0 (1-b) I_d^+$$
$$- \omega_0 b I_d^- - X^+ e^{-\frac{\tau}{\mu_0}} \text{ with } X^+ = \frac{\omega_0 \cdot S_{TOA}}{2\pi} b(\mu_0) \quad (A5)$$

$$-\overline{\mu} \frac{dI_d^-}{d\tau} = I_d^- - \omega_0 (1-b) I_d^- - \omega_0 b I_d^+$$
$$- X^- e^{-\frac{\tau}{\mu_0}} \text{ with } X^- = \frac{\omega_0 \cdot S_{TOA}}{2\pi} (1 - b(\mu_0)) \quad (A6)$$

where $\overline{\mu}$ is an average value of $\mu$ taken equal to $1/\sqrt{3}$ in the case of a beam source.

The backscattering coefficients $b(\mu)$ and $b$ are defined as:

$$b(\mu) = \frac{1}{2} \int_0^1 p(-\mu', \mu) d\mu' \text{ and } b = \frac{1}{2} \int_0^1 b(\mu) d\mu. \quad (A7)$$

Different choices of the phase function $p$ allow us to analytically solve these equations. An efficient definition is the Legendre polynomial expansion of the Henyey–Greenstein phase function. For simplicity, we stop at the *second* term here:

$$b(\mu) = \frac{1}{2}(1 - 3g \mu\overline{\mu}) \text{ and } b = \frac{1}{2}(1 - 3g \overline{\mu}^2). \quad (A8)$$

A more precise study using more terms in the Legendre polynomial expansion would require a numerical resolution of the equations. However, such a detailed analysis is not necessary for our study, where consequential approximations are already done with the 2D representation and on the lake description.

The equations are solved using the following boundary conditions:

$$I_d^+(\tau_0) = \alpha \left( I_d^-(\tau_0) + \frac{\mu_0 S_{TOA}}{2\pi\overline{\mu}} e^{-\tau_0/\mu_0} \right) \quad (A9)$$

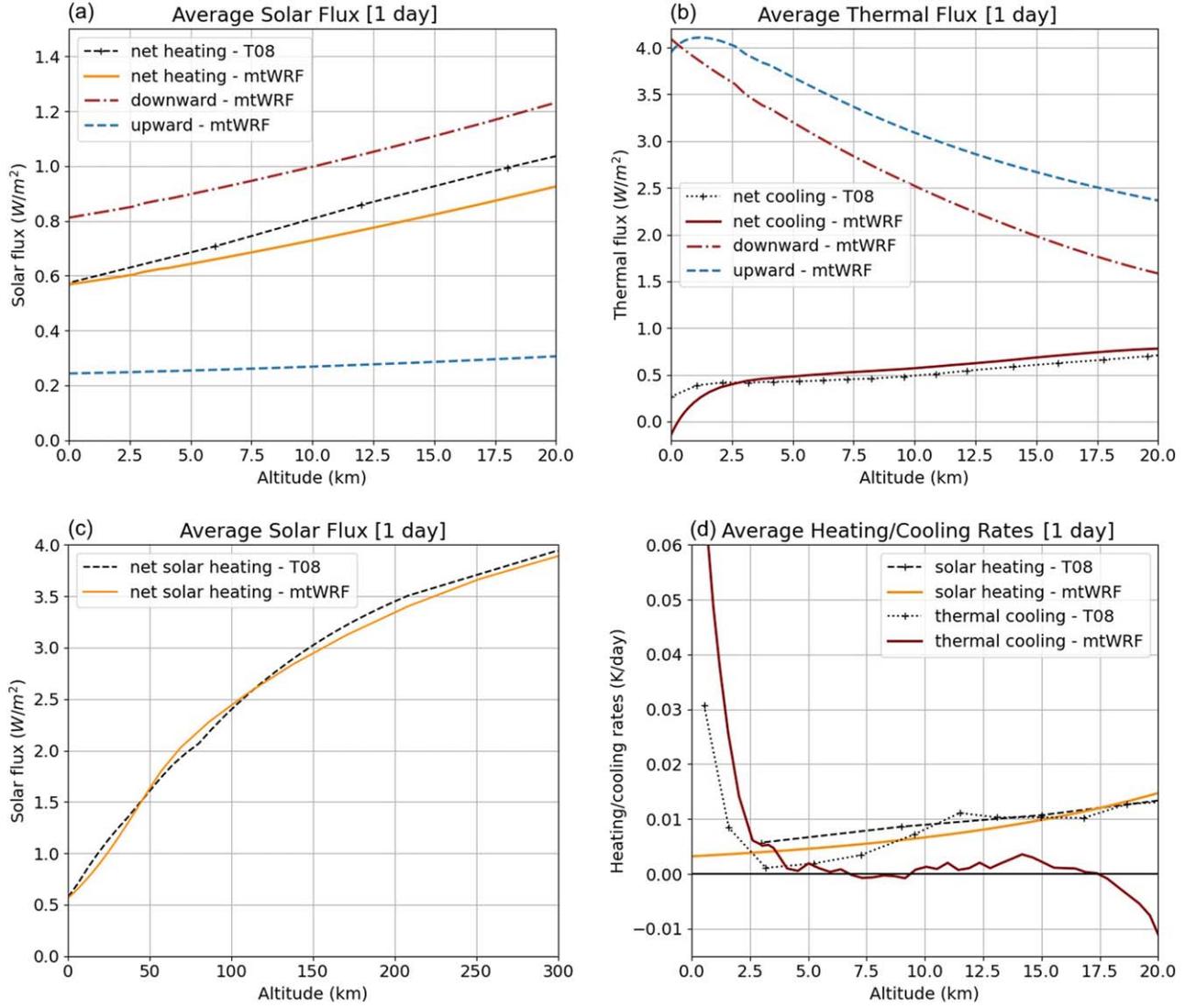

**Figure A1.** Daily average of the solar (panels (a) and (c)) and thermal (panel (b)) net fluxes, and the average heating/cooling rates (panel (d)) at the Huygens landing site. Fluxes in mtWRF (in a simulation in Huygens landing conditions, without lakes) are compared to fluxes retrieved from Huygens/DISR data by Tomasko et al. (2008a, T08).

$$I_d^-(\tau = 0) = 0 \quad (A10)$$

where $\alpha$ is the surface albedo, taken to be 0.3 on the land, and 0.1 on the lake (Tokano 2005b; Schneider et al. 2012).

The total flux is then obtained by:

$$F^+(\tau) = 2\pi\overline{\mu}\, I_d^+(\tau) \quad (A11)$$

$$F^-(\tau) = 2\pi\overline{\mu}\, I_d^-(\tau) + \mu_0\, S_{TOA}\, e^{-\tau/\mu_0}. \quad (A12)$$

### A.3. Emission and Absorption of the Thermal Radiation

The infrared flux is emitted by the surface and the atmosphere, and absorbed in the atmosphere. This scheme is taken from Schneider et al. (2012).

The absorption is implemented with the extinction optical depth:

$$\tau_{IR}(z) = \tau_{IR}^{surf}\left(f\,\frac{P(z)}{P_0} + (1-f)\left(\frac{P(z)}{P_0}\right)^2\right) \quad (A13)$$

where $\tau_{IR}^{surf} = 10$ is the surface optical depth, $P_0$ is the surface pressure, and $f = 0.15$ is a weight coefficient used to represent the absorption by a well-mixed absorber (linear term), and collision-induced absorption (quadratic term).

The infrared emission in the atmosphere and at the surface is obtained from the atmosphere and surface temperatures using the Stefan–Boltzmann law:

$$B_{atm}(\tau_{IR}) = \sigma_{SB}\, T_{atm}(\tau_{IR})^4$$
$$\text{and } B_{surf} = \epsilon\sigma_{SB}\, T_{surf}^4 \quad (A14)$$

where $\sigma_{SB}$ is the Stefan–Boltzmann constant and $\epsilon = 0.9$ is the surface emissivity.

The upward and downward thermal fluxes are described by the following equations:

$$\frac{dF_{IR}^+}{d\tau_{IR}} = F_{IR}^+(\tau_{IR}) - B_{atm}(\tau_{IR})$$
$$=> F_{IR}^+(\tau_{IR}) = F_{IR}^+(\tau_{IR} + d\tau_{IR})\, e^{-d\tau_{IR}}$$
$$+ B_{atm}(\tau_{IR})\,(1 - e^{-d\tau_{IR}}) \quad (A15)$$

$$-\frac{dF^-_{\text{IR}}}{d\tau_{\text{IR}}} = F^-_{\text{IR}}(\tau_{\text{IR}}) - B_{\text{atm}}(\tau_{\text{IR}})$$

$$=> F^-_{\text{IR}}(\tau_{\text{IR}} + d\tau_{\text{IR}}) = F^-_{\text{IR}}(\tau_{\text{IR}}) \times e^{-d\tau_{\text{IR}}} + B_{\text{atm}}(\tau_{\text{IR}})\,(1 - e^{-d\tau_{\text{IR}}}) \quad (\text{A16})$$

The upward flux is integrated upward with the boundary condition $F^+_{\text{IR}}(\tau^{\text{surf}}_{\text{IR}}) = B_{\text{surf}}$, and the downward flux is integrated downward with the boundary condition $F^-_{\text{IR}}(0) = 0$.

### A.4. Comparison to Tomasko et al. (2008a)

Figure A1 compares the daily averaged solar and thermal fluxes obtained with mtWRF to the net fluxes retrieved from Huygens/DISR data from Tomasko et al. (2008a).

As for the solar case, the values of the single scattering albedo $\omega_0$, the surface optical depth $\tau_0$, and the asymmetry factor $g$ are adjusted to fit these data, giving $\omega_0 = 0.9$, $\tau_0 = 8$, and $g = 0.85$. These values are coherent with previous estimations (Toon et al. 1992; Tomasko et al. 2008b). However, they are certainly modified compared to actual Titan values due to the approximations done in Appendix A.2. We were particularly careful to get the good surface flux, as it has the strongest impact on the surface heating and therefore on the air-surface exchanges and the local circulation.

To the contrary of the analytically resolved solar flux, the thermal flux is integrated over the altitude. With the top of the model being at 20 km, we adjusted the downward thermal flux at 20 km (= 1.55 W m$^{-2}$) to fit, at best, the average net flux and cooling rate given by Tomasko et al. (2008a). We note that Figure A1(b) shows a slight net downward thermal flux in the last hundreds of meters above the surface, opposite of Tomasko's lowest flux data point, which is slightly upward. However, Tomasko et al. (2008a) warned about the uncertainties in the lowest-altitude data points of the thermal flux. Given the lack of more precise measurements, we kept our model as is. Our model starts to drop out on the upper levels, but this has no impact on our results, since lake-induced effects are only visible in the last kilometers above the surface.

### ORCID iDs

Audrey Chatain https://orcid.org/0000-0002-2252-3254

Alejandro Soto https://orcid.org/0000-0002-2333-0307

Ricardo Hueso https://orcid.org/0000-0003-0169-123X